\documentclass[a4paper,fleqn,usenatbib]{mnras}

\usepackage{newtxtext,newtxmath}

\usepackage[T1]{fontenc}
\usepackage{ae,aecompl}

\usepackage{graphicx}	
\usepackage{amsmath}	
\usepackage{amssymb}	
\usepackage{array}
\usepackage{makecell}
\newcolumntype{C}[1]{>{\centering\arraybackslash}p{#1}}
\usepackage{caption}

\title[Triggering radio-intermediate HERGs]{Do AGN triggering mechanisms vary with radio power? I. Optical morphologies of radio-intermediate HERGs}

\author[J. C. S. Pierce et al.]{J. C. S. Pierce,$^{1}$\thanks{E-mail: jcspierce1@sheffield.ac.uk}
C. N. Tadhunter,$^{1}$
C. Ramos Almeida,$^{2,3}$
P. S. Bessiere$^{4}$
\newauthor and M. Rose$^{1}$
\\
\\
$^{1}$Department of Physics and Astronomy, University of Sheffield, Sheffield S3 7RH, UK\\
$^{2}$Instituto de Astrof\'{i}sica de Canarias, Calle V\'{i}a L\'{a}ctea, s/n, E-38205 La Laguna, Tenerife, Spain\\
$^{3}$Departamento de Astrof\'{i}sica, Universidad de La Laguna, E-38205 La Laguna, Tenerife, Spain\\
$^{4}$Instituto de Astrof\'{i}sica, Facultad de F\'{i}sica, Pontificia Universidad Cat\'{o}lica de Chile, Casilla 306, Santiago 22, Chile
}

\date{Accepted XXX. Received YYY; in original form ZZZ}

\pubyear{2018}

\begin{document}
\label{firstpage}
\pagerange{\pageref{firstpage}--\pageref{lastpage}}
\maketitle

\begin{abstract}
Radio AGNs with intermediate radio powers are capable of driving multi-phase outflows in galaxy bulges, and are also more common than their high-radio-power counterparts. In-depth characterisation of the typical host galaxies and likely triggering mechanisms for this population is therefore required in order to better understand the role of radio AGN feedback in galaxy evolution. Here, we use deep optical imaging data to study the detailed host morphologies of a complete sample of 30 local radio AGNs with high-excitation optical emission (HERG) spectra and intermediate radio powers ($z < 0.1$; 22.5 $<$ log(L$_{\rm 1.4GHz}$) $< 24.0$ W\,Hz$^{-1}$). The fraction of hosts with morphological signatures of mergers and interactions is greatly reduced compared to the 2 Jy radio-powerful galaxies (log(L$_{\rm 1.4GHz}$) $> 25.0$ W\,Hz$^{-1}$) with strong optical emission lines: 53 $\pm$ 9 per cent and 94 $\pm$ 4 per cent, respectively. In addition, the most radio-powerful half of the sample has a higher frequency of morphological disturbance than the least radio-powerful half (67 $\pm$ 12 per cent and 40 $\pm$ 13 per cent, respectively), including the eight most highly-disturbed galaxies. This suggests that the importance of triggering nuclear activity in HERGs through mergers and interactions reduces with radio power. Both visual inspection and detailed light profile modelling reveal a mixed population of early-type and late-type morphologies, contrary to the massive elliptical galaxy hosts of radio-powerful AGNs. The prevalence of late-type hosts could suggest that triggering via secular, disk-based processes has increased importance for HERGs with lower radio powers (e.g. disk instabilities, large scale bars).
\end{abstract}




\section{Introduction}
\label{sec:int}

The feedback effects caused by powerful radio jets from active galactic nuclei (AGNs) have rendered them an important component of current models of galaxy evolution. On the scales of individual galaxies, these jets are seen to drive massive outflows of gas in multiple phases \citep[e.g.][]{mor05a,mor05b,nes06,holt08,nes10,rus17}, potentially having a significant influence on star formation in the inner regions of the galaxy and regulating the fuelling of the AGN.

In many radio galaxies, these jets are seen to inflate larger-scale cavities in the hot interstellar/intracluster gas (ISM/ICM) that can propagate well beyond the extent of the hosts \citep[see][for a review]{mn07}, also giving their feedback effects importance in a cosmological framework. Since the bright central galaxies in high density galaxy clusters are predominantly radio AGNs \citep[e.g.][]{best07}, they are commonly invoked as the solution to the `cooling flow problem' associated with these environments \citep{fab94}, providing an injection of energy that can counteract the cooling of hot intracluster gas via X-ray emission. This would also restrict the build up of mass in these brightest cluster galaxies (BCGs). Thus, radio AGNs are often considered key to accounting for the lack of observed galaxies at the upper end of galaxy luminosity or stellar mass functions, relative to theoretical predictions \citep[e.g.][]{bow06,vog14,cro06,cro16}.

The heating of the hot ISM/ICM by relativistic jets is commonly implemented into cosmological simulations through the inclusion of `radio-mode' AGN activity. \citet[][]{best06} have demonstrated that this type of feedback is dominated by relatively low radio power sources (log(L$_{\rm 1.4GHz}) \lesssim 24.0$ W\,Hz$^{-1}$), which often have weak or `low-excitation' optical emission lines and are labelled as low-excitation radio galaxies (LERGs). Such radio AGNs are thought to be fuelled by radiatively-inefficient accretion flows of hot gas that lead to low black hole accretion rates \citep[e.g.][]{hard07,bh12}.

In contrast, `quasar mode' AGN activity is commonly assumed to drive broad winds that can heat and expel the warm and cool phases of the ISM in the bulges of the host galaxies. This mode is associated with the radiatively-efficient accretion of cold gas at high black hole accretion rates, resulting in high-excitation optical emission-line spectra. This wind-based feedback may therefore be important for high-excitation radio galaxies (HERGs), which are believed to be fuelled in this way \citep[e.g.][]{bh12}. However, such objects also possess powerful radio jets that are themselves able to heat and expel the ISM in the host galaxies.

Indeed, there is now strong evidence to suggest that the radio jets of HERGs with intermediate radio powers (22.5 $<$ log(L$_{\rm 1.4GHz}$) $< 25.0$ W\,Hz$^{-1}$) are capable of causing kinematic disturbances in the multi-phase gas of their host galaxies \citep[][]{tad14a,har15,ram17,vm17}, perhaps even more significantly than those with higher radio powers \citep[][]{mul13}. In addition, it is evident from the local radio galaxy luminosity function that radio-intermediate AGNs are considerably more common than their counterparts at higher radio powers \citep[][]{bh12}. Detailed characterisation of the population of radio-intermediate HERGs is therefore crucial for investigating the role of radio AGN feedback in galaxy evolution.

In order to correctly implement radio-intermediate HERG feedback into models of galaxy evolution, it is important to know what types of host galaxies they are associated with. At high radio powers, the general population of radio AGN hosts is predominantly composed of elliptical galaxies with high stellar masses \citep{mms64,dun03,best05b,tad11}. It is seen, however, that in samples of local radio galaxies selected at high flux densities (e.g. 3CR and 2 Jy), the few that show signs of disk-like components in optical images are typically found towards lower radio powers, and half lie in the radio-intermediate regime \citep[][]{tad16}. In addition, it appears that the frequency of late-type hosts could increase at lower radio powers $-$ \citet[][]{sad14} find that 30 per cent of their radio AGN hosts are spiral galaxies at low redshifts (mostly L$\rm _{1.4GHz}$\,$\sim$ $10^{22-23}$\,W\,Hz$^{-1}$). Furthermore, Seyfert galaxies, arguably the counterparts of HERGs at the lowest radio powers, are predominantly associated with late-type morphologies \citep[e.g.][]{ada77}. Therefore, the radio-intermediate objects are crucial for investigating the possible transition in host galaxy morphologies along the HERG sequence.

In addition, it is important to determine the likely triggering mechanisms for the nuclear activity. Previous deep optical imaging studies of large samples of powerful radio galaxies demonstrate that those with strong optical emission lines commonly show morphological signatures of galaxy mergers at relatively high surface brightnesses \citep{heck86,sh89a,sh89b}. Most recently, \citet[][]{ram11} have shown that 94 $\pm$ 4 per cent of the strong-line radio galaxies (SLRGs)\footnote{Strong-line radio galaxies are selected based on [OIII] emission-line equivalent width, but show a strong overlap with the HERG population \citep[see discussion in][]{tad16}.} in the 2 Jy sample exhibit such tidal features. Comparison with images of control samples of elliptical galaxies matched in absolute magnitude and surface brightness depth also reveals an excess of high-surface-brightness tidal features for the 2 Jy SLRGs relative to the non-active population \citep{ram12}. The SLRGs also show a preference for group-like environments \citep{ram13}, which are suitable for frequent galaxy-galaxy collisions without the high relative galaxy velocites found in clusters that might reduce the merger rate \citep{quinn84,pb06}. This, amongst other evidence, has led to suggestions that the cold gas necessary for radiatively efficient accretion in HERGs is supplied by mergers and interactions, in line with optically selected AGNs \citep{cro06,but10,bh12,hb14,min14}.

In contrast to the situation for the high radio power HERGs selected in samples such as the 3CR and 2 Jy, the lack of systematic study of radio-intermediate HERGs means that the typical host galaxies and fuelling mechanisms have not yet been determined for this population. Here, we present the first in a series of papers which use deep optical imaging of local HERGs with intermediate radio powers to investigate the host morphologies and likely triggering mechanisms for their nuclear activity. We use deep images of a complete sample of 30 radio-intermediate HERGs to analyse their detailed optical morphologies, and determine the host galaxy types from both visual inspection and light profile modelling. In \S\ref{sec:sam_obs_red}, we detail the selection of our sample and describe the observations and subsequent data reduction. The analysis of the reduced images and subsequent results are outlined in \S\ref{sec:analysis}. Our results are discussed in \S\ref{sec:disc}, and \S\ref{sec:conc} summarises our work and concludes the study. A cosmology described by $H_{0} = 73.0$ km\,s$^{-1}$\,Mpc$^{-1}$, $\Omega_{\rm m} = 0.27$ and $\Omega_{\rm \Lambda} = 0.73$ is assumed throughout this paper.

\section{Sample selection, Observations and Reduction}
\label{sec:sam_obs_red}

\subsection{Sample selection}
\label{sec:sample_sel}

The sample was selected from the catalog of 18,286 local radio AGNs produced by \citet{bh12}, which comprises all AGN host galaxies within SDSS DR7 \citep[][]{aba09} with counterpart radio emission at 1.4\,GHz in the FIRST \citep[][]{bec95} and/or NVSS \citep[][]{con98} surveys. The objects were classified as AGNs based on their SDSS optical emission spectra, which were also used to classify the AGNs as either high-excitation or low-excitation radio galaxies (HERGs/LERGs). This latter division is based on either `excitation index' (EI) from \citet{but10}, diagnostic diagrams from \citet{kew06}, [OIII] emission line equivalent width, or plots of [NII]/H$\alpha$ versus [OIII]/H$\alpha$ from \citet{cid10} $-$ see \citet{bh12} for further details on the classification. Note that this selection differs from the strong-line/weak-line radio galaxy (SLRG/WLRG) division $-$ based solely on the equivalent width
of the [OIII]$\lambda$5007 emission line $-$ used in other studies \citep{tad98,ram11,ram12,ram13}, as is discussed in \citet{tad16}.

To construct the sample, all AGNs in the catalog with intermediate radio powers in the range 22.5 $<$ log(L$_{\rm 1.4GHz}$) $< 24.0$ W\,Hz$^{-1}$ and high-excitation emission spectra (HERGs) were selected. In almost all cases, the radio powers were derived from the measured NVSS fluxes for the targets, to ensure that as little as possible of the radio emission was lost due to resolution effects. The NVSS fluxes could not be used for the targets J0911 and J1358, due to contamination by other radio sources within the resolution limit of the NVSS survey, and so the measured FIRST fluxes were used in these cases. The sample was also restricted to low redshifts ($z < 0.1$) to ensure sensitivity to any low surface brightness tidal features that may be present. The right ascension ($\alpha$) values for the targets were then restricted to the range 07h 15m $< \alpha <$ 16h 45m, such that the targets could be easily observed in the same observing window. The final sample consists of 30 objects, and provides a large, representative sample of local HERGs with intermediate radio powers that is complete within these constraints $-$ see Table~\ref{tab:target_info} for further information on the individual targets. Although we cannot entirely rule out the contamination of the sample by radio-quiet AGNs in star forming galaxies, we consider this to be unlikely based on the analysis of far-infrared data; see Appendix~\ref{app:SFcont}.

\begin{figure}
\centering
    \includegraphics[width=\columnwidth]{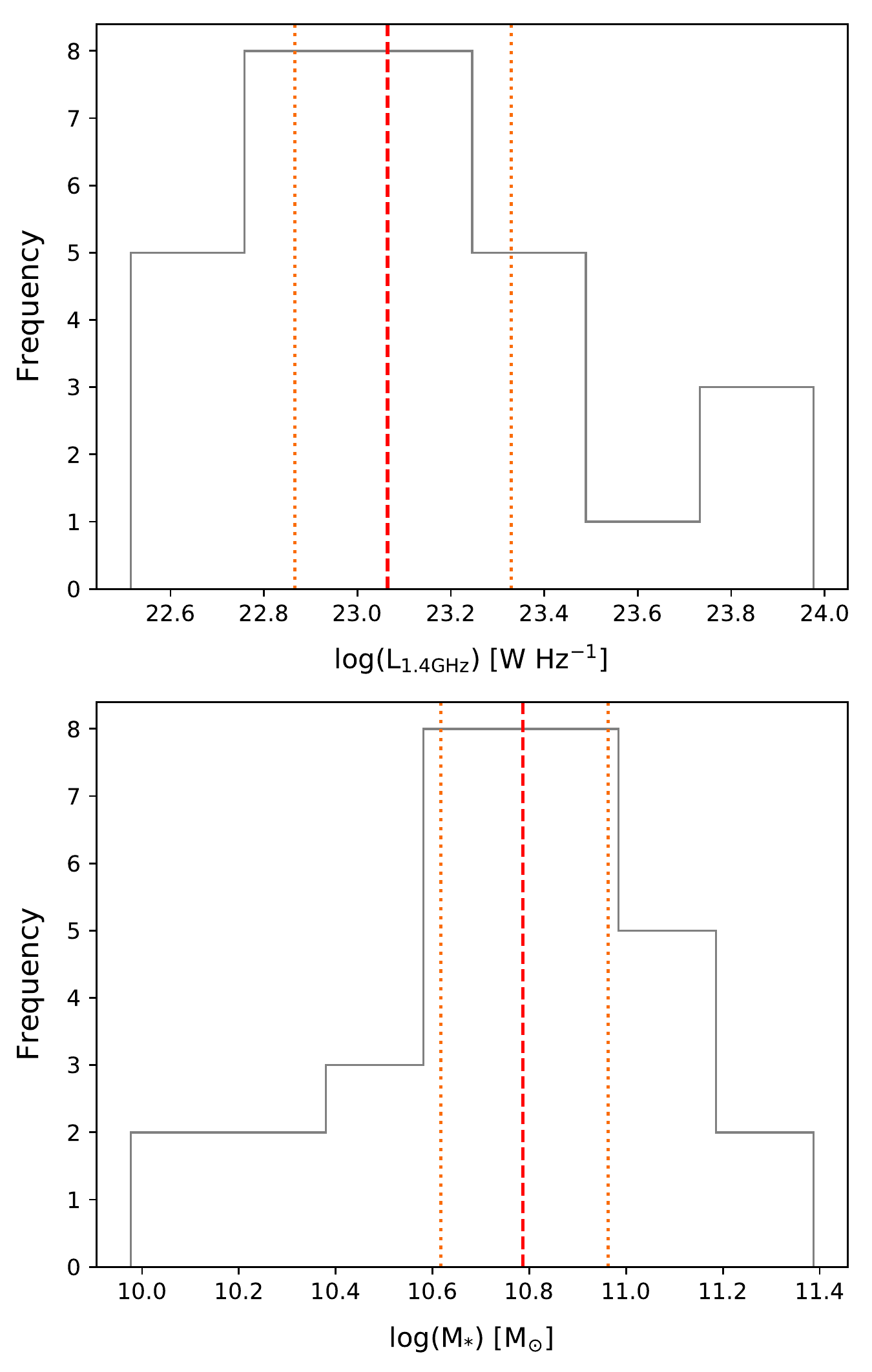}
    \caption{The L$\rm_{1.4GHz}$ radio power (upper panel) and stellar mass (lower panel) distributions for the galaxies in the sample. The median values are indicated by the red dashed lines, whilst the 25th and 75th percentiles are displayed as orange dotted lines.}
   \label{fig:hists}
\end{figure}

The radio power and stellar mass distributions for the HERGs in the sample are displayed in Figure~\ref{fig:hists}, with their 25th, 50th (median) and 75th percentiles indicated. The galaxies have stellar masses in the range 10.0 $<$ log($\rm M_{*}/M_{\odot}$) $< 11.4$ though are skewed towards higher masses, with a median of 10$\rm ^{10.8}$ $\rm M_{\odot}$. The radio power distribution is slightly skewed towards the lower radio luminosities in the range, with a median of 10$^{23.06}$ W\,Hz$^{-1}$, as expected from the local luminosity function.

\begin{table*}
	\centering
	\caption{Basic information for all 30 targets within the observed sample, selected as detailed in \S\ref{sec:sample_sel}. Column 1 lists the full SDSS IDs for the targets, with their in-text abbreviations included in brackets. Columns 2, 3 and 4 give the spectroscopic redshifts, \emph{r}-band extinction (in magnitudes) and \emph{r}-band magnitudes (corrected for extinction) for the targets, as obtained through the SDSS DR7 Object Explorer tool (available at: http://cas.sdss.org/dr7/en/tools/explore/). The \emph{r}-band extinction values are calculated from the full-sky 100\,$\mu$m maps of \citet{sch98}, whilst the raw \emph{r}-band magnitudes considered are are those given by the best fitting model between standard de Vaucouleurs and exponential profiles. Column 5 denotes the radio luminosities of the galaxies at 1.4\,GHz derived from their NRAO VLA Sky Survey (NVSS) fluxes \citep{con98}, or Faint Images of the Radio Sky at Twenty-cm (FIRST) fluxes \citep{bec95} in the cases of J0911 and J1358 (see \S\ref{sec:sample_sel}). Columns 6 and 7 provide the [OIII]\,$\rm \lambda5007$ luminosities and stellar masses taken from the MPA/JHU value-added catalogue for SDSS DR7 (available at: https://wwwmpa.mpa-garching.mpg.de/SDSS/DR7/). Columns 8 and 9 show the date and the atmospheric seeing for the target observations, respectively.}
	\label{tab:target_info}
	\begin{tabular}{lcccccccc}
		\hline
		\\
		SDSS ID (Abbr.) & $z$ & \makecell{$A_{\rm r}$\\(mag)} & \makecell{SDSS \emph{r} mag\\(mag)} & \makecell{log(L$_{\rm 1.4GHz}$)\\(W\,Hz$^{-1}$)} & \makecell{log(L$_{\rm [OIII]}$)\\(W)} & \makecell{log(M$_{*}$)\\(M$_{\odot}$)} & Obs. date & \makecell{Seeing \small{FWHM}\\(arcsec)}\\
		\\
		\hline
		\\
   J072528.47$+$434332.4	(J0725)	&	0.069	&	0.25	&	16.37	&	23.06	&	33.90	&	10.9	&	2017-03-29	&	1.30\\
  J075756.72$+$395936.1	(J0757)	&	0.066	&	0.14	&	15.36	&	23.98	&	34.12	&	10.8	&	2017-03-28	&	1.31\\
  J081040.29$+$481233.1	(J0810)	&	0.077	&	0.14	&	15.87	&	23.70	&	33.54	&	10.9	&	2017-03-28	&	1.34\\
  J082717.80$+$125430.0	(J0827)	&	0.065	&	0.10	&	16.00	&	22.71	&	33.36	&	10.7	&	2017-03-30	&	1.38\\
  J083637.83$+$440109.5	(J0836)	&	0.055	&	0.08	&	15.25	&	23.94	&	33.79	&	11.1	&	2017-03-28	&	1.24\\
  J083856.90$+$261037.5	(J0838)	&	0.051	&	0.14	&	16.41	&	22.62	&	33.57	&	10.3	&	2017-03-30	&	1.37\\
  J090239.51$+$521114.7	(J0902)	&	0.098	&	0.05	&	15.50	&	23.94	&	33.25	&	11.4	&	2017-03-29	&	1.21\\
J091107.04$+$454322.6	(J0911)	&	0.098	&	0.05	&	17.33	&	22.62	&	33.26	&	10.7	&	2017-03-29	&	1.24\\
J093141.45$+$474209.0	(J0931)	&	0.049	&	0.04	&	17.07	&	22.54	&	33.38	&	10.0	&	2017-03-31	&	1.40\\
J095058.69$+$375758.8	(J0950)	&	0.041	&	0.05	&	15.29	&	23.37	&	33.26	&	10.8	&	2017-03-29	&	1.21\\
J103655.60$+$380321.4	(J1036)	&	0.051	&	0.06	&	14.45	&	22.82	&	33.79	&	11.2	&	2017-03-31	&	1.41\\
J110009.47$+$100312.1	(J1100)	&	0.064	&	0.08	&	17.24	&	22.85	&	32.98	&	10.2	&	2017-03-31	&	1.40\\
J110852.61$+$510225.6	(J1108)	&	0.070	&	0.04	&	15.89	&	23.07	&	33.84	&	11.0	&	2017-03-31	&	1.57\\
J114740.08$+$334720.4	(J1147)	&	0.031	&	0.05	&	14.29	&	22.52	&	33.94	&	10.7	&	2017-03-31	&	1.68\\
J115020.82$+$010423.1	(J1150)	&	0.078	&	0.07	&	17.70	&	22.97	&	33.29	&	10.7	&	2017-03-30	&	1.94\\
J120611.42$+$350525.6	(J12061)	&	0.081	&	0.05	&	15.71	&	22.92	&	33.89	&	11.1	&	2017-03-30	&	1.40\\
J120640.31$+$104652.8	(J12064)	&	0.089	&	0.06	&	17.62	&	23.41	&	33.02	&	11.0	&	2017-03-28	&	1.24\\
J123641.09$+$403225.7	(J1236)	&	0.096	&	0.06	&	17.82	&	23.08	&	33.57	&	10.5	&	2017-03-29	&	1.14\\
J124322.55$+$373858.0	(J1243)	&	0.086	&	0.05	&	15.69	&	23.35	&	33.97	&	11.2	&	2017-03-28	&	1.14\\
J125739.56$+$510351.4	(J1257)	&	0.097	&	0.03	&	16.34	&	23.20	&	33.33	&	11.1	&	2017-03-29	&	1.14\\
J132450.59$+$175815.0	(J1324)	&	0.086	&	0.06	&	16.67	&	23.35	&	33.36	&	10.7	&	2017-03-28	&	1.07\\
J135152.97$+$465026.1	(J1351)	&	0.096	&	0.04	&	17.04	&	23.26	&	33.21	&	10.6	&	2017-03-28	&	1.08\\
J135817.73$+$171236.7	(J1358)	&	0.095	&	0.10	&	16.87	&	23.21	&	33.93	&	10.9	&	2017-03-28	&	1.27\\
J141217.69$+$242735.9	(J1412)	&	0.069	&	0.06	&	15.42	&	23.19	&	33.79	&	11.1	&	2017-03-29	&	1.67\\
J152948.01$+$025536.7	(J1529)	&	0.077	&	0.13	&	16.56	&	22.93	&	33.21	&	10.8	&	2017-03-30	&	1.54\\
J155511.00$+$270056.2	(J1555)	&	0.083	&	0.12	&	17.36	&	22.92	&	33.47	&	10.4	&	2017-03-31	&	2.05\\
J160110.88$+$431138.1	(J1601)	&	0.072	&	0.04	&	17.09	&	22.99	&	34.24	&	10.4	&	2017-03-31	&	1.41\\
J160953.46$+$133147.8	(J1609)	&	0.036	&	0.11	&	15.06	&	23.01	&	32.86	&	10.8	&	2017-03-31	&	1.60\\
J162237.28$+$075349.5	(J1622)	&	0.088	&	0.22	&	17.10	&	23.19	&	33.77	&	10.6	&	2017-03-29	&	1.40\\
J163045.51$+$125752.7	(J1630)	&	0.065	&	0.15	&	16.93	&	22.78	&	33.89	&	10.1	&	2017-03-31	&	2.02\\
		   \\   
		\hline
	\end{tabular}
\end{table*}

\subsection{Observations}
\label{sec:obs}

The deep optical imaging data were obtained on four consecutive nights in 2017 March, using the Wide-Field Camera (WFC) attached to the 2.5m Isaac Newton Telescope (INT) at the Observatorio del Roque de los Muchachos, La Palma. The seeing full width at half maximum (FWHM) ranged between 1.07 and 2.05 arcsec over the course of the observations, with a median of 1.38 arcsec. The individual seeing values for the observations were calculated by the \texttt{THELI} package for astronomical image reduction and processing \citep[see][for details]{sch13}, using plots of half-light radius against magnitude for foreground stars in the final coadded images (post-reduction). The first night of observations was photometric, but intermittent thin cloud rendered the remaining nights not entirely photometric.
Details on the individual target observations are presented in Table~\ref{tab:target_info}.

The WFC consists of four thinned, anti-reflection-coated 2048 $\times$ 4100 pixel CCDs separated by gaps of $660 - 1098$ \textrm{\micron}. With pixel sizes of 0.333 arcsec\,pixel$^{-1}$, this provides a large total field-of-view of order 34 $\times$ 34 arcmin$^{2}$. The images were taken through the WFC Sloan \emph{r}-band filter ($\rm{\lambda_{eff}} = 6240$\,\AA, $\Delta \lambda = 1347$\,\AA), chosen to be consistent with existing SDSS \emph{r'}-band observations of our targets \citep[see][for further details]{fuk96} and the GMOS-S \emph{r'}-band ($z<0.4$) and \emph{i'}-band ($0.4<z<0.7$) observations of radio-loud galaxies in the 2 Jy sample performed by \citet[][]{ram11}. Each target was observed using 4 $\times$ 700s exposures, yielding total exposure times of 2800s per target. Such exposure lengths were necessary for the detection of low-surface-brightness tidal features, but were divided into four separate exposures to ensure that the target galaxies were not saturated in individual images. A square dithering pattern was employed to overcome the gaps in the images introduced by the spacings between the CCDs, using four pointing offsets of 30 arcsec. Flat-field and image defect corrections are also improved by this process.

\subsection{Reduction}
\label{sec:red}

\begin{figure*}
    \centering
    \vspace{0.3cm}
    \includegraphics[scale=1.17]{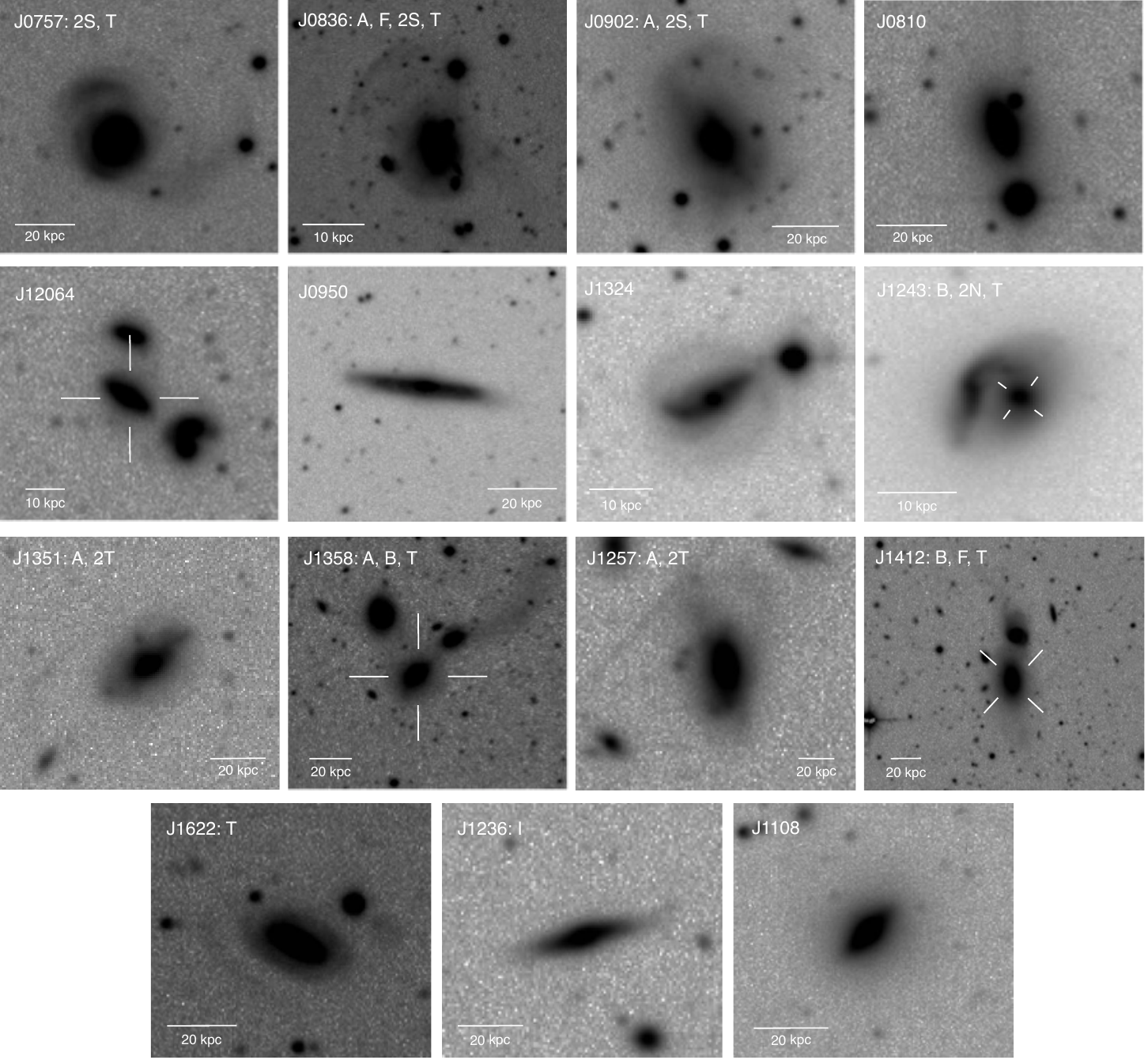}
    \vspace{0.2cm}
    \caption{Deep \emph{r}-band images of the half of the HERGs in the sample with the highest radio powers, lying within the range 23.06 $<$ log(L$_{\rm 1.4GHz}$) $< 24.0$ W\,Hz$^{-1}$. The galaxies are also ordered in decreasing radio power from top-left to bottom-right within the figure. The images are centred on the targets and are identified within crosses on the images, where ambiguous. The abbreviated SDSS IDs are shown for each target, along with any interaction signatures from the classification scheme outlined in \S\ref{sec:class} that are associated with the hosts or companions clearly involved in the interactions. Approximate scale bars (in kpc) are displayed in all cases. The images are oriented such that north is up and east is left.}
    \label{fig:RP_top}
\end{figure*}

\begin{figure*}
    \centering
    \vspace{0.3cm}
    \includegraphics[scale=1.17]{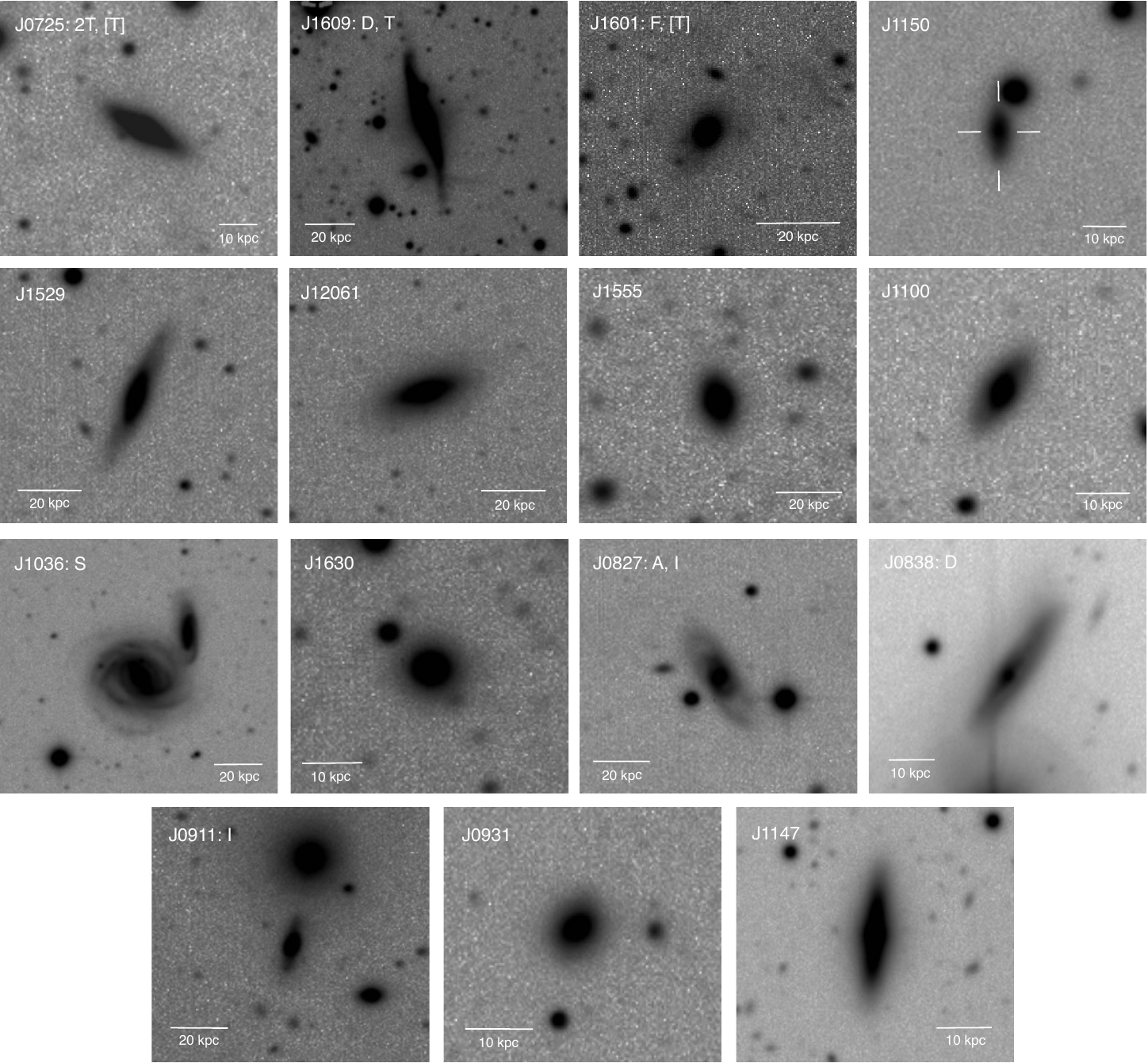}
    \vspace{0.2cm}
    \caption{As Figure~\ref{fig:RP_top}, but for the half of the HERGs in the sample with the lowest radio powers, lying within the range 22.5 $<$ log(L$_{\rm 1.4GHz}$) $< 23.06$ W\,Hz$^{-1}$.}
    \label{fig:RP_bottom}
\end{figure*}

The subsequent image processing, from initial reduction to construction of the final mosaic images, was done using \texttt{THELI} \citep[][]{sch13}. Bias corrections for each observation were performed via the subtraction of master bias frames, whilst master flat-field frames were used for flat-field corrections. These master frames were constructed by median-combining individual calibration frames that were taken on the same night as the given observation. The individual flat-field images were dome flats with exposure times $\sim$ 3s. Removal of the overscan level was performed on all calibration and science frames.

An astrometric solution for all calibrated images was calculated in \texttt{THELI} through cross-matching object catalogues produced for each image by \texttt{SExtractor} \citep[][]{ber96} with the all-sky USNO-B1 catalog \citep[][]{mon03}. The comparison of these catalogues and the generation of the astrometric solution was performed within \texttt{THELI} using the \texttt{Scamp} pipeline for astrometry and photometry \citep[see][]{ber06}. Cross-matching was only carried out for sources with apparent magnitudes $\leq$\,20 to prevent the astrometry from being overconstrained, whilst still providing a sufficiently large number of objects to obtain a good solution. 

To correct for variations in the sky background level across the images, \texttt{THELI} was used to produce a sky model and subtract it from the calibrated data. Prior to model creation, objects with a minimum of five connected pixels at a value of 1.5\,$\sigma$ above the measured background level were detected and removed using \texttt{SExtractor}. A sky model convolved with a Gaussian kernel with a FWHM of 60 pixels was found to produce the best results in almost all cases; the exception was J0902, for which a FWHM of 200 pixels was required due to a large gradient in the background level near a very bright foreground star. Coaddition of the four separate calibrated images was performed to produce the final mosaic image for each target, using the positional information from the prior astrometric solutions.

To calibrate the images photometrically, \texttt{THELI} was used to calculate photometric zero points for the coadded images of each of the targets. This was performed through comparison of the derived instrumental magnitudes for stars in the image field with their catalogued SDSS magnitudes. Only stars with instrumental magnitude errors $\leq 0.05$ (as derived by \texttt{SExtractor}) and maximum counts $\leq 70$ per cent of the saturation level were used for this process. This method has the advantage of automatically correcting for photometric variability throughout the nights, since the calibration stars are observed at the same time and at the same position on the sky as the galaxy targets. It also removes the reliance on average zero points derived from standard star observations.

\section{Data analysis and results}
\label{sec:analysis}

Following the reduction outlined above (\S\ref{sec:red}), the final coadded images were analysed to investigate the detailed optical morphologies of the AGN host galaxies in our radio-intermediate sample. The first goal of this was to identify and characterise any tidal features and other interaction signatures associated with the galaxies, in order to provide an indication of the importance of merger-based triggering of the nuclear activity (see \S\ref{sec:tf}). Following this, the morphological classes of the host galaxies were classified via both visual inspection (\S\ref{sec:vis}) and detailed modelling of their surface brightness profiles using the two-dimensional fitting algorithm \texttt{GALFIT} \citep[][see \S\ref{sec:galfit}]{peng02,peng10}. Determining whether the population is primarily composed of early- or late-type galaxies provides an initial impression of how they compare to other radio galaxy samples, and is a factor that could be linked to the dominant triggering mechanism for the nuclear activity. The bulge-disk decomposition available from the \texttt{GALFIT} models also allows quantitative estimates of the level of disk- or bulge-dominance in the light profiles to be made, assisting with the morphological classifications. The details of this data analysis are presented here.

\subsection{Tidal features}
\label{sec:tf}

\subsubsection{Classification}
\label{sec:class}
Visual inspection of the images was performed by five human classifiers in order to identify any interaction signatures linked with the AGN host galaxies. To be consistent with the analysis performed for the deep images of the 2 Jy radio galaxies \citep{ram11,ram12,ram13}, the classification scheme of \citet{ram11} was used for this process. This scheme is adapted from that produced by \citet{heck86} for their study of a selection of powerful radio galaxies, and is based on the morphological disturbances caused by tidal forces in the galaxy merger simulations of \citet{tt72} and \citet{quinn84}. The tidal features and other interaction signatures are classified as follows:
\begin{enumerate}
\item \textbf{Tail (T)} $-$ a narrow curvilinear feature with roughly radial orientation;
\item \textbf{Fan (F)} $-$ a structure similar to a tail, but that is shorter and broader;
\item \textbf{Shell (S)} $-$ a curving filamentary structure with a roughly tangential orientation relative to a radial vector from the main body of the galaxy;
\item \textbf{Bridge (B)} $-$ a feature that connects a radio galaxy with a companion;
\item \textbf{Amorphous halo (A)} $-$ the galaxy halo is misshapen in an unusual way in the image;
\item \textbf{Irregular (I)} $-$ the galaxy is clearly disturbed, but not in a way that fits into any of the previous classifications;
\item \textbf{Multiple nuclei (e.g. 2N, 3N)} $-$ two or more brightness peaks within a distance of 9.6\,kpc;
\item \textbf{Dust lane (D)} $-$ a clear linear dark structure within the galaxy.
\end{enumerate}

A galaxy with a morphology that did not meet any of these criteria, or that only showed evidence of dust lanes (viii), was classified as undisturbed and hence non-interacting. Examples of each of the different classifications are indicated in Figures~\ref{fig:RP_top} and~\ref{fig:RP_bottom}, which display the deep \emph{r}-band images for the halves of the sample with highest and lowest radio powers, respectively $-$ images within each Figure are also ordered by decreasing radio power from top-left to bottom-right.

Each classifier was provided with a FITS image of each of the radio galaxies in the sample and asked to identify and classify any interaction signatures that they believed to be associated with them. A brief description of their characteristics (e.g. orientation, extent, faintness) was also required to help avoid confusion when comparing classifications from different individuals. A qualitative estimate of the confidence of the classifications was also recorded by each assessor to aid with the process of combining results to produce final classifications. This method was carried out by each individual independently to avoid any influence from the results of others. 

In order to avoid biases introduced by the galaxy properties, most importantly by their radio powers, only the cosmological scale (kpc\,arcsec$^{-1}$) for each target was provided to each classifier prior to image inspection $-$ this value was required to identify targets that host multiple nuclei. The independent results were combined by considering each individual classification as a `vote', such that those voted for by the majority of the classifiers were considered in the final classifications for each target. Failing a favoured result, the noted certainty provided for each individual classification was considered. If no final decision had been made, the classifications were discussed by the assessors to gain an overall consensus. The final classifications for the interaction signatures associated with all galaxies in the sample are indicated in Table~\ref{tab:SB_info}.

\begin{table}
\caption{The number and corresponding proportion of galaxies deemed as having clear interaction signatures and no clear interaction signatures. For those with clear interaction signatures, specific proportions for each of the classifications outlined in \S\ref{sec:class} are also presented, in order of decreasing frequency. Only one occurrence of a certain classification of signature is considered for each galaxy, but each different classification is considered individually. The listed uncertainties are standard binomial errors.}
\label{tab:prop}
\begin{tabular}{m{4cm}C{1.4cm}C{1.4cm}}
\hline \\
Classification               & Number of galaxies & Proportion of sample \\
 \\ \hline
                \\
Clear interaction signatures    & 16                 & 53 $\pm$ 9 \%    \\
No clear interaction signatures & 14                 & 47 $\pm$ 9 \%        \\
\\
Tidal tail (T)                  & 12                 & 40 $\pm$ 9 \%       \\
Amorphous halo (A)              & 6                  & 20 $\pm$ 7 \%        \\
Shell (S)                       & 4                  & 13 $\pm$ 6 \%        \\
Irregular (I)                   & 3                  & 10 $\pm$ 5 \%         \\
Fan (F)                         & 3                  & 10 $\pm$ 5 \%         \\
Bridge (B)                      & 2                  & 7 $\pm$ 5 \%         \\
Dust lane (D)                   & 2                  & 7 $\pm$ 5 \%         \\
Double nucleus (2N)             & 1                  & 3 $\pm$ 3 \%        \\
\\
\hline
\end{tabular}
\end{table}

\begin{figure*}
\centering
    \includegraphics[width=14cm]{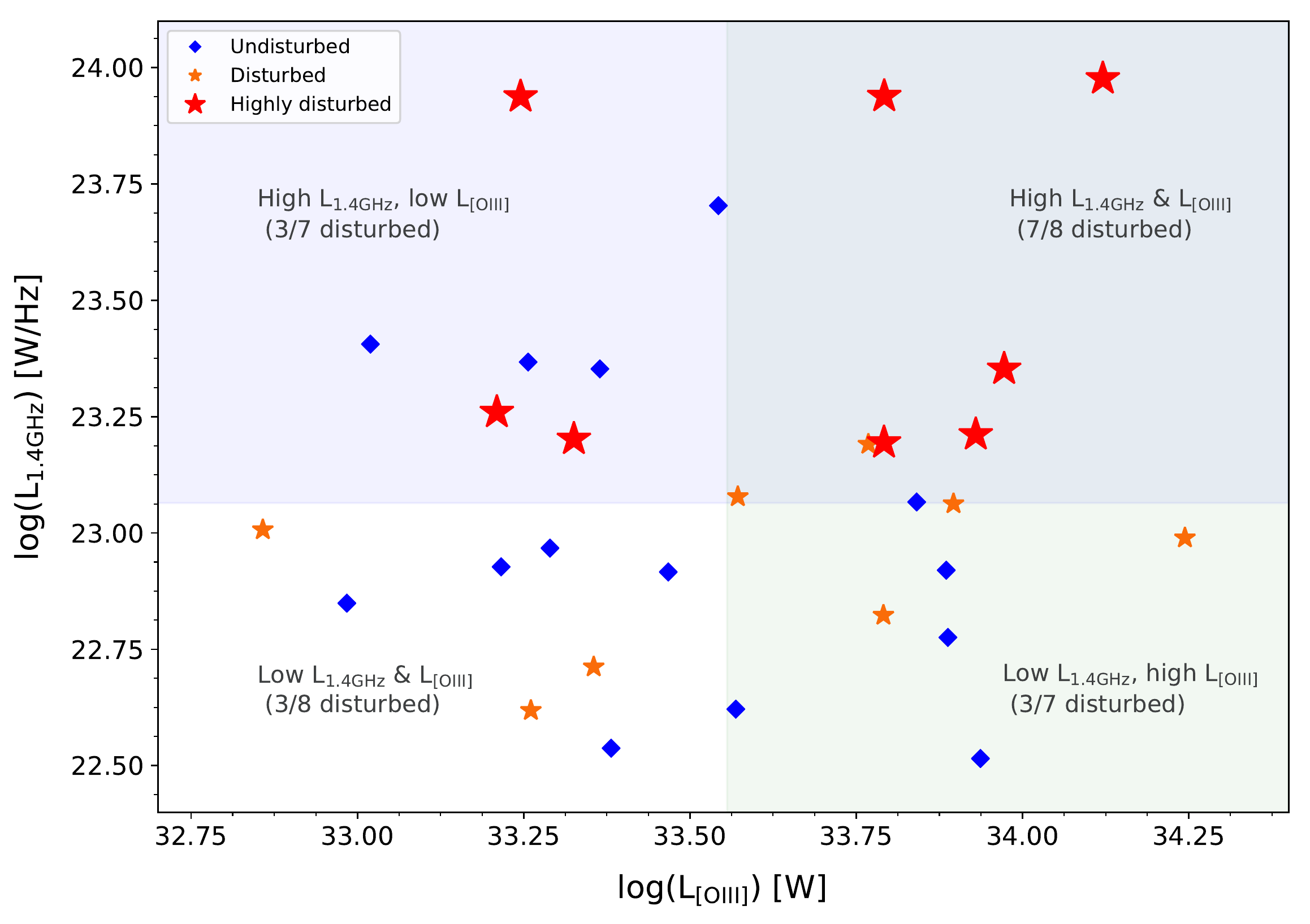}
    \caption{A plot of the radio powers at 1.4 GHz against the [OIII]\,$\lambda$5007 luminosities for the galaxies in the sample. Galaxies with morphological signatures of disturbance are plotted as orange stars, and the larger red stars indicate those with high levels of disturbance. Undisturbed galaxies are plotted as blue diamonds. The blue shaded region indicates the half of the sample with the highest radio powers whilst the green-grey shaded region denotes the half with the highest [OIII] luminosities. The proportion of disturbed galaxies in each quadrant is indicated.}
   \label{fig:RPvsOIII}
\end{figure*}

\subsubsection{Proportions}
\label{sec:prop}

Once the classification process had been carried out, the proportions of galaxies in the sample that showed signs of interaction were analysed. In an attempt to reduce ambiguity, the classified galaxies were separated into two categories: (i) `clear interaction signatures' and (ii) `no clear sign of interaction'. Category (i) only contained galaxies with interaction signatures that the assessors had deemed as certain following the visual classification process. The remaining galaxies, with indefinite or non-existent interaction signatures, were placed in category (ii). Due to the nature of this division, the number of galaxies that are classified as interacting can be considered to be a lower limit. Dust lanes (the D classification) were not considered to be a clear interaction signature. The proportion of galaxies in the sample that were placed in categories (i) and (ii) are presented in Table~\ref{tab:prop}, along with the proportion that showed evidence for each individual type of interaction signature (i.e., T, F, S, B, A, I, 2N, D). 

Overall, 53 $\pm$ 9 per cent of the sample showed some signature of a past or on-going interaction. When the sample is divided into two equally populated bins in radio power, it is found that galaxies in the most radio-powerful half of the sample (23.06 $<$ log($\rm L_{1.4GHz}$) $< 24.0$ W\,Hz$^{−1}$) show interaction signatures more frequently than those in the least radio-powerful half (22.5 $<$ log($\rm L_{1.4GHz}$) $< 23.06$ W\,Hz$^{−1}$) $-$ 67 $\pm$ 12 per cent and 40 $\pm$ 13 per cent, respectively (see Figure~\ref{fig:TF_prop}). After conducting a two-proportion Z-test, it is found that the null hypothesis that these two proportions of interacting galaxies are the same can be rejected at a confidence level of 85.7 per cent (1.5\,$\sigma$). However, given the proportions measured, it is found that a minimum of 63 galaxies would be needed in each half of the sample for the Z-test to reject the null hypothesis at the 3\,$\sigma$ level, illustrating the need for more data. An improved statistical analysis will be performed in future papers in this series when further deep imaging data has been reduced and analysed.

Interestingly, the exact same proportions are found when dividing the sample into two equally populated bins in [OIII]\,$\lambda$5007 luminosity: 67 $\pm$ 12 per cent for the half with the highest luminosities (33.56 $<$ log($\rm L_{\rm [OIII]}$) $< 34.24$ W) and 40 $\pm$ 13 per cent for the half with the lowest luminosities (32.86 $<$ log($\rm L_{\rm [OIII]}$) $< 33.56$ W). Plotting L$\rm _{1.4 GHz}$ against L$\rm _{[OIII]}$ for both the morphologically disturbed and undisturbed galaxies in the sample (Figure~\ref{fig:RPvsOIII}) reveals no significant correlation between the two parameters, however, as confirmed by a Pearson correlation test ($r=0.077$, $p=0.685$). This is in line with what is seen at radio-intermediate powers in other studies \citep[e.g.][]{best05b}. As such, L$\rm _{1.4 GHz}$ and L$\rm _{[OIII]}$ are independent of one another within the sample, and, in fact, it can be seen in Figure~\ref{fig:RPvsOIII} that the two sets of proportions are not provided by the same galaxies. 

Following cursory visual inspection, however, eight of the galaxies in the sample show clear, highly disturbed morphologies or are involved in interactions with large-scale tidal tails, suggestive of major interactions: J0757, J0836, J0902, J1243, J1257, J1351, J1358 and J1412 (large red stars in Figure~\ref{fig:RPvsOIII}). All of these galaxies lie in the half of the sample with the highest radio powers, and hence none lie in the half with the lowest radio powers. Using a two-proportion Z-test, it is found that the null hypothesis that the proportion of highly disturbed galaxies in these two halves of the sample are the same can be rejected at a confidence level of 99.9 per cent (3.3\,$\sigma$). We note that given that there were only eight of such galaxies in the entire sample, a more significant result could not have been observed.

Only 5 out of these 8 galaxies lie in the half with the highest [OIII]\,$\lambda$5007 luminosities, however, with the remaining 3 out of 8 seen to be in the half with the lowest [OIII]\,$\lambda$5007 luminosities. In this case, a two-proportion Z-test reveals that the null hypothesis that these two proportions are the same can only be rejected at a confidence level of 59.1 per cent (0.8\,$\sigma$). These results suggest that the highly disturbed galaxy morphologies are linked with the radio powers of the AGNs, but are unlikely to be linked with their [OIII]\,$\lambda$5007 luminosities. We reserve further discussion on this matter for \S\ref{sec:maj_int}.

\subsubsection{Surface brightness}
\label{sec:sb}

\begin{figure}
\centering
    \includegraphics[width=8cm]{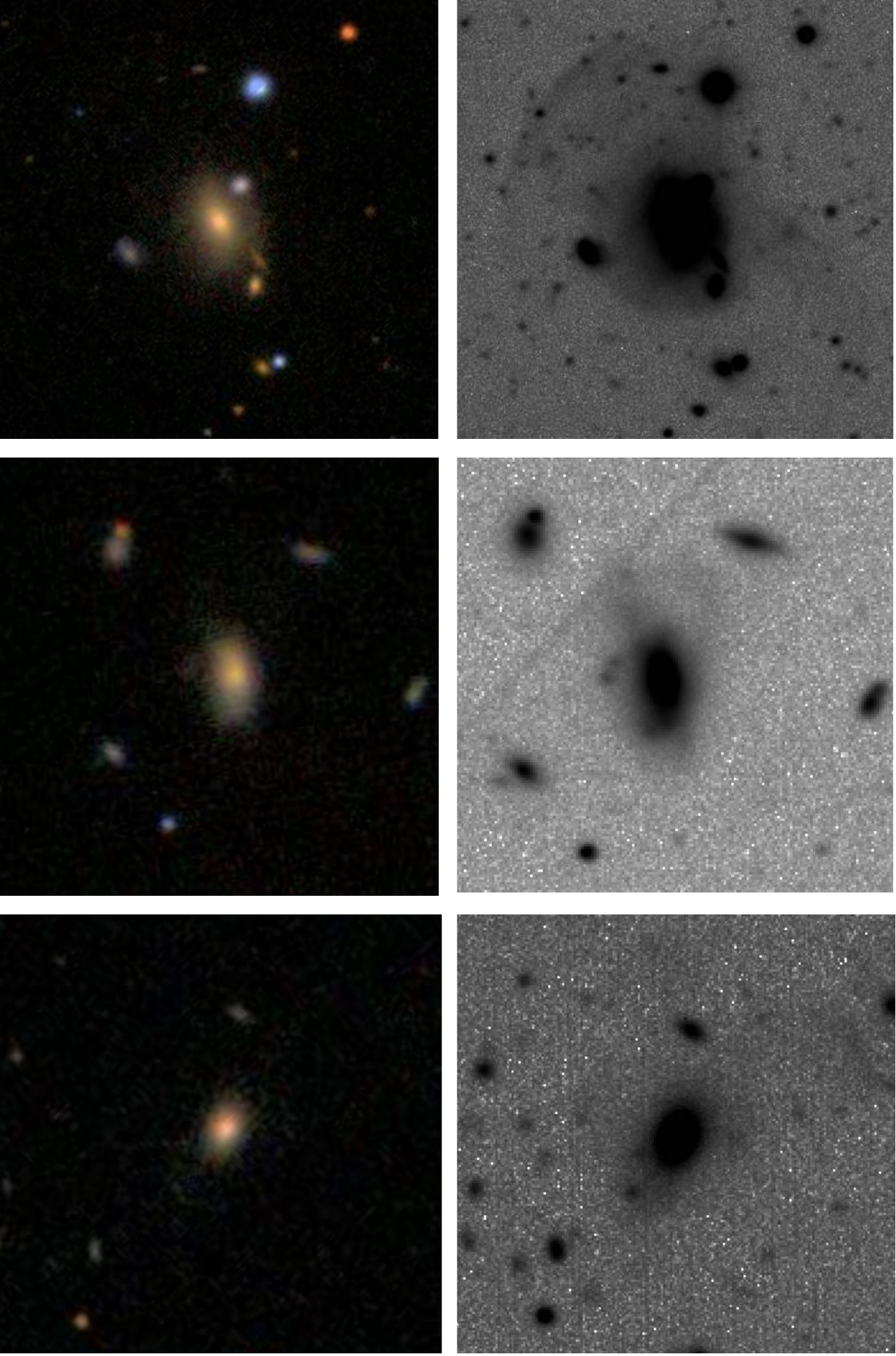}
    \caption{Example SDSS (left column; composite \emph{g-r-i}) and INT/WFC (right column; \emph{r}-band) images for three galaxies in the sample. The INT/WFC observations reveal interaction signatures that are not visible in the SDSS imaging, highlighting the importance of deep imaging observations for the detection of low surface brightness tidal features.}
   \label{fig:SDSS_comp}
\end{figure}

\begin{table*}
\centering
\caption{Surface brightness information for the targets in the sample. Column 1 displays the target names, which are the abbreviated forms of their SDSS IDs, as in Table~\ref{tab:target_info}. Column 2 denotes the level of surface brightness dimming in magnitudes per unit area as provided by the NASA/IPAC Extragalactic Database (NED). Columns 3 and 4 give the classification and measured surface brightness for the tidal features associated with each target. Column 5 shows these surface brightnesses after corrections for surface brightness dimming and foreground galactic extinction ($A_r$ in Table~\ref{tab:target_info}), and cosmological \emph{K}-correction following \citet[][online calculator available at http://kcor.sai.msu.ru/]{chil10}. The surface brightness limits derived from $\sigma_{\rm sky}$ and $3\sigma_{\rm sky}$, as per \S\ref{sec:sb} are provided in Columns 6 and 7, respectively. All surface brightness measurements are in mag\,arcsec$^{-2}$. The values associated with tidal features with measured surface brightnesses fainter than the $3\sigma_{\rm sky}$ surface brightness limit but brighter than the $\sigma_{\rm sky}$ limit are indicated in square brackets.}
\label{tab:SB_info}
\begin{tabular}{ccccccc}
\hline
\\
Target & Dimming & Morphology & \makecell{\large{$\mu$}$_{\rm AB}$\\\small{(mag\,arcsec$^{-2}$)}}  & \makecell{\large{$\mu$}$_{\rm AB}^{\rm corr}$\\\small{(mag\,arcsec$^{-2}$)}}  & \makecell{\large{$\mu$}$_{\rm AB}^{\rm \sigma_{\rm sky}}$\\\small{(mag\,arcsec$^{-2}$)}}  & \makecell{\large{$\mu$}$_{\rm AB}^{\rm 3\sigma_{\rm sky}}$\\\small{(mag\,arcsec$^{-2}$)}}  \\
\\
\hline
\\
J0725  & 0.292      & 2T, {[}T{]}    & 26.10, 26.50, {[}26.85{]} & 25.49, 25.88, {[}26.24{]}  & 27.83         & 26.63         \\
J0757  & 0.279      & 2S, T      & 23.23, 23.56, 25.41 & 22.75, 23.08, 24.93 & 27.99         & 26.80         \\
J0810  & 0.326      & -          & -                   & -                   & 28.39         & 27.20         \\
J0827  & 0.279      & A, I       & -                   & -                   & 27.40         & 26.21         \\
J0836  & 0.237      & A, F, 2S, T   & 22.64, 25.25, 26.08, 26.36 & 22.28, 24.89, 25.71, 25.99 & 28.28         & 27.09         \\
J0838  & 0.220        & D          & -                   & -                   & 27.93         & 26.74         \\
J0902  & 0.409      & A, 2S, T   & 23.47, 26.38, 23.60 & 22.92, 25.83, 23.05 & 28.20         & 27.00         \\
J0911  & 0.409      & I          & -                   & -                   & 27.89         & 26.70         \\
J0931  & 0.209      & -          & -                   & -                   & 28.16         & 26.97         \\
J0950  & 0.176       & -          & -                   & -                   & 28.00         & 26.81         \\
J1036  & 0.218      & S          & 25.10               & 24.78               & 28.23         & 27.03         \\
J1100  & 0.274      & -          & -                   & -                   & 27.97         & 26.77         \\
J1108  & 0.295      & -          & -                   & -                   & 27.99         & 26.80         \\
J1147  & 0.136      & -          & -                   & -                   & 28.27         & 27.08         \\
J1150  & 0.332      & -          & -                   & -                   & 27.41         & 26.21         \\
J12061 & 0.341      & -          & -                   & -                   & 27.20         & 26.01         \\
J12064 & 0.376      & -          & -                   & -                   & 28.18         & 26.98         \\
J1236  & 0.401      & I          & -                   & -                   & 28.09         & 26.89         \\
J1243  & 0.361      & B, 2N, T   & 20.62, 21.66  & 20.12, 21.66       & 27.72         & 26.53         \\
J1257  & 0.402      & A, 2T      & 24.81, 25.97        & 24.30, 25.46        & 27.83         & 26.64         \\
J1324  & 0.362      & -          & -                   & -                   & 27.89         & 26.69         \\
J1351  & 0.402      & A, 2T      & 23.15, 26.50        & 22.64, 25.98        & 28.25         & 27.06         \\
J1358  & 0.399      & A, B, T       & 25.43, 26.47     & 24.84, 25.88    & 28.12         & 26.92         \\
J1412  & 0.293      & B, F, T       & 24.12, 24.23, 25.63  & 23.69, 23.80, 25.21    & 28.07         & 26.88         \\
J1529  & 0.322      & -          & -                   & -                   & 27.78         & 26.59         \\
J1555  & 0.346      & -          & -                   & -                   & 28.12         & 26.93         \\
J1601  & 0.301      & F, {[}T{]}     & 25.08, {[}26.96{]}      & 24.68, {[}26.56{]}      & 28.09         & 26.90         \\
J1609  & 0.152      & D, T       & 26.17               & 25.88               & 27.99         & 26.80         \\
J1622  & 0.367      & T          & 26.51               & 25.84               & 28.05         & 26.86         \\
J1630  & 0.273      & -          & -                   & -                   & 28.14         & 26.95        
\\
\\
\hline
\end{tabular}
\end{table*}

To determine the surface brightness sensitivity achieved by the observations, as is important for comparison with other studies, surface brightness limits were calculated for the final coadded images. As outlined by \citet{duc15}, determinations of surface brightness limits in deep imaging programs often vary, making effective comparisons difficult. Consequently, we clarify our technique here. 

Our method is based on measuring the variations in the sky background level in the final coadded images, and closely follows that of \citet{atk13} for their analysis of faint tidal features in wide-field images from the Canada-France-Hawaii Telescope Legacy Survey (CFHTLS). We stress that the measurements were performed \emph{after} all reduction, including the flattening achieved through the subtraction of a sky background model (see \S\ref{sec:red}). The total counts were measured for 40 apertures of one-arcsecond in diameter for each of the final images. These were distributed in suitable locations $-$ with minimal or no influence from the halos of bright objects $-$ across the field of the CCD in which the targets were centred (WFC CCD4). The standard deviation in the sky background level within the apertures ($\sigma_{\rm sky}$) was then determined and converted to an apparent magnitude to provide final limiting surface brightness measurements. The mean $\sigma_{\rm sky}$ and $3\sigma_{\rm sky}$ limiting surface brightnesses over all observations are 28.0 mag\,arcsec$^{-2}$ and 26.8 mag\,arcsec$^{-2}$, respectively. The measured standard deviation for these quantities is 0.3 mag\,arcsec$^{-2}$. $\sigma_{\rm sky}$ and $3\sigma_{\rm sky}$ limiting surface brightness magnitudes for all targets are presented in Table~\ref{tab:SB_info}.

Following this, the surface brightness of each of the detected tidal features was determined. For consistency, these measurements were carried out using a similar method to that used for the limiting surface brightnesses, using the average counts measured in one-arcsecond apertures evenly distributed over the full extent of the tidal feature. In this case, however, matching background apertures were also placed in suitable locations surrounding the tidal feature. This allowed the measured count level to be corrected for background contributions arising from either residual sky background or underlying smooth galaxy emission. 

The measured surface brightnesses of all visually-detected tidal features are listed along with their classifications in Table~\ref{tab:SB_info}. The measurements were corrected for surface brightness dimming using values provided by the NASA/IPAC Extragalactic Database (NED), and for extragalactic extinction using the \emph{r}-band extinction values listed in Table~\ref{tab:target_info}. Cosmological \emph{K}-corrections were also applied following the redshift-colour method of \citet{chil10}, using the integrated \emph{g}-\emph{r} colours calculated from the available SDSS DR7 magnitudes. These corrections 
were performed to be consistent with those applied by \citet{ram11} and ensure meaningful comparisons between the measured surface brightnesses of the two studies. The values calculated following all corrections are also presented in the table. Surface brightness measurements that are brighter than the $\sigma_{\rm sky}$ surface brightness limit for the target observation but fainter than the 3$\sigma_{\rm sky}$ limit are considered as uncertain detections, and are included inside square brackets in the table. The tidal features that were confidently detected ranged in surface brightness from 20.6 mag\,arcsec$^{-2}$ to 27.0 mag\,arcsec$^{-2}$, with a median of 25.4 mag\,arcsec$^{-2}$. The standard deviation of these measurements is 1.6 mag\,arcsec$^{-2}$. 

Due mainly to degeneracies in galaxy merger model parameters (e.g. age, mass ratio, impact angle), it is currently difficult to determine physically meaningful properties from the exact measured surface brightnesses of tidal features. As a result, no detailed conclusions are drawn from the values are reported here, although the values are useful in a relative sense for comparing between different studies and samples. For instance, here we note that $\sim$\,40 per cent of the tidal features that we detect are fainter than the limiting surface brightness of standard depth SDSS images \citep[$\sim$\,25 mag\,arcsec$^{-2}$;][]{dri16}, highlighting the importance of the improved depth provided by the targeted observations. This can be seen clearly for the examples shown in Figure~\ref{fig:SDSS_comp}, where the INT/WFC observations reveal interaction signatures that are not detected in SDSS images.

\subsection{Host morphologies}
\label{sec:hosts}

\subsubsection{Visual inspection}
\label{sec:vis}
When carrying out the visual inspection process outlined in \S\ref{sec:class}, the classifiers were also asked to provide comments on the general morphological properties of the host galaxies. This was primarily done to provide an impression of their morphological types. For late-type galaxies, however, assessors also noted the level of disk warping and the rough disk orientation (from edge-on to face-on), which could assist with the interpretation of the interaction signature classifications. Late-type galaxies were those seen to have a significant disk or spiral-like morphology (discarding lenticular galaxies), and were found to be the most popular category within the sample (43 per cent; 13/30). Early-type galaxies, appearing as elliptical-like or lenticular, were found to comprise 37 per cent (11/30) of the population, whilst the remaining 20 per cent (6/30) were judged to have morphologies too disturbed to classify $-$ the `merger' class. The visual classifications for all the galaxies in the sample are presented in the last column of Table~\ref{tab:galfit_info}.

Classifications based on visual inspection are also available for our sample in the public data release for the Galaxy Zoo project \citep{lin08,lin11}. The project required participants to classify galaxies in composite \emph{g}-, \emph{r}- and \emph{i}-band images from SDSS into one of the following categories: ``Elliptical (E)", ``ClockWise spirals (CW)", ``AntiClockWise spirals (ACW)", ``Edge-on spirals (Edge)", ``Don't Know (DK)" or ``Merger (MG)". The vote fractions provided for the E classification are taken to indicate the likelihood of a galaxy being of early-type, whereas the sum of the vote fractions for the CW, ACW and Edge classifications (``Combined Spiral", CS) are taken as the likelihood of a galaxy being of late-type. 

In our case, we only accepted a classification if the value of the lower error boundary on the vote fraction for the most popular category was greater than the value of the upper error boundary on the least popular category. In other words, the lower error boundary on the early-type vote fraction was required to be greater than the upper error boundary on the late-type vote fraction for a galaxy to be classed as early-type, and vice versa. The error boundaries considered for the vote fractions were 95 per cent confidence intervals. The vote fraction of the most popular classification was also required to be greater than 0.5 to ensure that the other two categories (DK and MG) were not favoured. Should these conditions not be met, the galaxy morphology was classed as uncertain. From this, it is found that 30 per cent of the sample (9/30) are classed as early-types, 37 per cent (13/30) as late-types, and the remaining 33 per cent (10/30) as uncertain. 

As part of the Galaxy Zoo project, an attempt was also made to correct for biases introduced by the magnitude limit of SDSS and varying galaxy appearance with redshift $-$ for details, see \citet{bam09} and \citet{lin11}. `Debiased' vote fractions are available for 28 of the galaxies in the sample, and when these results are considered, it is found that 25 per cent (7/28) are classed as early-types, 43 per cent (12/28) as late-types and the remaining 32 per cent (9/28) as uncertain. Since the biases tend to lead to overestimates of the number of early-type galaxies relative to late-types, debiasing of the vote fractions tends to increase the vote fractions for late-type galaxies relative to early-types on the whole, as is reflected here. The visual classification results from both the authors and the Galaxy Zoo project are summarised in Table~\ref{tab:host_prop} and discussed in \S\ref{sec:disc_hosts}.

\subsubsection{Light profile modelling}
\label{sec:galfit}

\begin{table*}
\centering
\caption{The results from detailed modelling of the host galaxy light profiles using \texttt{GALFIT}. The abbreviated target names are shown in Column 1, as listed in Table~\ref{tab:target_info}. For each model component, Columns 2-6 provide: (2) the component type (i.e. S\'{e}rsic or PSF) and its classification as bulge-like (B), disk-like (D) or intermediate (I) for multi-component fits; (3) the S\'{e}rsic index; (4) the effective radius in kpc; (5) the axis ratio; (6) the derived apparent magnitude. Columns 7-10 then present information derived from the overall models (combination of all components): (7) total apparent magnitude; (8) ratio of bulge-to-disk light (where relevant); (9) ratio of bulge-to-total light; (10) ratio of disk-to-total light. The morphological types from the visual inspection performed by the authors are also presented in Column 11, with their classifications as either mergers or early- (E) or late-type (L) galaxies indicated in brackets.}
\label{tab:galfit_info}
\begin{tabular}{C{1cm}C{1.5cm}C{0.8cm}C{1cm}C{0.8cm}C{1cm}C{1cm}C{0.7cm}C{0.7cm}C{0.7cm}C{2.3cm}}
\hline
&               &       &           &            &           &            &            &          &           &           \\
Target & Comp. Type     & $n$ & R$\rm _e$ (kpc) & $\rm b/a$ & $m_{c}$ & $m_{t}$ & B/D  & B/T  & D/T & Visual class.    \\
    (1)          &   (2)    &    (3)       &   (4)         &   (5)        &    (6)        &     (7)       &    (8)      &    (9)       &     (10)   &   (11)   \\
&               &       &           &            &           &            &            &          &           &           \\
\hline
&               &       &           &            &           &            &            &          &           &           \\
J0725  & S\'{e}rsic  (B) & 2.00 & 0.78            & 0.67      & 17.94   & 16.75   & 0.50 & 0.33 & 0.67 & Lenticular (E)   \\
       & S\'{e}rsic  (D) & 0.92 & 5.06            & 0.22      & 17.19   &         &      &      &      &                  \\
J0757  & -              & -    & -               & -         & -       & -       & -    & -    & -    & Merger           \\
J0810  & -              & -    & -               & -         & -       & -       & -    & -    & -    & Lenticular (E)   \\
J0827  & S\'{e}rsic  (B) & 2.00 & 1.42            & 0.58      & 17.73   & 16.35   & 0.39 & 0.28 & 0.72 & Spiral (L)       \\
       & S\'{e}rsic  (D) & 0.75 & 6.16            & 0.46      & 16.70   &         &      &      &      &                  \\
J0836  & S\'{e}rsic      & 5.00 & 6.70            & 0.64      & 15.08   & 15.08   & -    & -    & -    & Merger           \\
J0838  & S\'{e}rsic  (B) & 3.29 & 4.02            & 0.35      & 16.08   & 15.20   & 0.79 & 0.44 & 0.56 & Edge-on disk (L) \\
       & S\'{e}rsic  (D) & 0.50 & 9.67            & 0.22      & 15.83   &         &      &      &      &                  \\
J0902  & S\'{e}rsic      & 5.00 & 14.42           & 0.70      & 15.23   & 15.23   & -    & -    & -    & Merger           \\
J0911 & S\'{e}rsic (I)     & 2.00 & 1.74            & 0.46      & 18.38   & 17.52   & -    & -    & -    & Spiral (L)       \\
       & S\'{e}rsic (I)     & 1.50 & 6.79            & 0.25      & 18.21   &         &      &      &      &                  \\
       & PSF            & -    & -               & -         & 21.91   &         &      &      &      &                  \\
J0931  & S\'{e}rsic      & 3.13 & 0.55            & 0.36      & 17.82   & 17.02   & -    & -    & -    & Elliptical (E)   \\
       & S\'{e}rsic      & 2.00 & 2.55            & 0.89      & 17.72   &         &      &      &      &                  \\
J0950  & S\'{e}rsic  (B) & 2.54 & 6.74            & 0.23      & 15.78   & 15.12   & 1.21 & 0.55 & 0.45 & Edge-on disk (L) \\
       & S\'{e}rsic  (D) & 0.50 & 14.06           & 0.11      & 15.99   &         &      &      &      &                  \\
J1036  & S\'{e}rsic  (B) & 5.00 & 6.22            & 0.54      & 14.94   & 14.36   & 1.41 & 0.59 & 0.41 & Spiral (L)       \\
       & S\'{e}rsic  (D) & 0.50 & 12.63           & 0.76      & 15.31   &         &      &      &      &                  \\
J1100  & S\'{e}rsic  (B) & 2.00 & 1.38            & 0.34      & 17.89   & 17.41   & 1.80 & 0.64 & 0.36 & Spiral (L)       \\
       & S\'{e}rsic  (D) & 0.52 & 4.49            & 0.39      & 18.53   &         &      &      &      &                  \\
J1108  & S\'{e}rsic  (B) & 2.00 & 1.50            & 0.74      & 16.65   & 15.87   & 1.48 & 0.60 & 0.40 & Lenticular (E)   \\
       & S\'{e}rsic  (D) & 1.36 & 4.86            & 0.20      & 16.86   &         &      &      &      &                  \\
       & PSF (B)        & -    & -               & -         & 18.28   &         &      &      &      &                  \\
J1147  & -              & -    & -               & -         & -       & -       & -    & -    & -    & Lenticular (E)   \\
J1150  & S\'{e}rsic  (B) & 2.11 & 2.64            & 0.52      & 18.51   & 17.69   & 0.89 & 0.47 & 0.53 & Spiral (L)       \\
       & S\'{e}rsic  (D) & 0.50 & 4.08            & 0.20      & 18.38   &         &      &      &      &                  \\
J12061 & S\'{e}rsic  (B) & 5.00 & 10.00           & 0.40      & 15.77   & 15.62   & 6.78 & 0.87 & 0.13 & Lenticular (E)   \\
       & S\'{e}rsic  (D) & 0.96 & 5.75            & 0.20      & 17.85   &         &      &      &      &                  \\
J12064 & S\'{e}rsic  (B) & 5.00 & 5.82            & 0.62      & 17.78   & 17.35   & 2.04 & 0.67 & 0.33 & Spiral (L)       \\
       & S\'{e}rsic  (D) & 0.50 & 4.49            & 0.18      & 18.55   &         &      &      &      &                  \\
J1236  & S\'{e}rsic  (B) & 2.00 & 1.33            & 0.62      & 19.74   & 17.88   & 0.22 & 0.18 & 0.82 & Edge-on disk (L) \\
       & S\'{e}rsic  (D) & 0.61 & 6.29            & 0.16      & 18.09   &         &      &      &      &                  \\
J1243  & -              & -    & -               & -         & -       & -       & -    & -    & -    & Merger           \\
J1257  & -              & -    & -               & -         & -       & -       & -    & -    & -    & Merger           \\
J1324  & -              & -    & -               & -         & -       & -       & -    & -    & -    & Spiral (L)       \\
J1351  & -              & -    & -               & -         & -       & -       & -    & -    & -    & Merger           \\
J1358  & S\'{e}rsic      & 3.79 & 1.63            & 0.27      & 18.71   & 16.96   & -    & -    & -    & Elliptical (E)   \\
       & S\'{e}rsic      & 2.91 & 5.11            & 0.61      & 17.20   &         &      &      &      &                  \\
J1412  & S\'{e}rsic      & 4.79 & 2.56            & 0.56      & 16.18   & 15.38   & -    & -    & -    & Elliptical (E)   \\
       & S\'{e}rsic      & 2.00 & 9.66            & 0.62      & 16.08   &         &      &      &      &                  \\
J1529  & S\'{e}rsic  (B) & 5.00 & 2.78            & 0.70      & 18.05   & 16.56   & 0.34 & 0.25 & 0.75 & Edge-on disk (L) \\
       & S\'{e}rsic  (D) & 0.50 & 5.63            & 0.19      & 17.84   &         &      &      &      &                  \\
       & S\'{e}rsic  (D) & 0.82 & 10.78           & 0.18      & 17.46   &         &      &      &      &                  \\
&               &       &           &            &           &            &            &          &           &           \\
       \hline
\end{tabular}
\end{table*}

\begin{table*}
\contcaption{}
\begin{tabular}{C{1cm}C{1.5cm}C{0.8cm}C{1cm}C{0.8cm}C{1cm}C{1cm}C{0.7cm}C{0.7cm}C{0.7cm}C{2.3cm}}
\hline
&               &       &           &            &           &            &            &          &           &           \\
Target & Comp. Type     & $n$ & R$\rm _e$ (kpc) & $\rm b/a$ & $m_{c}$ & $m_{t}$ & B/D  & B/T  & D/T & Visual class.    \\
    (1)          &   (2)    &    (3)       &   (4)         &   (5)        &    (6)        &     (7)       &    (8)      &    (9)       &     (10)   &   (11)   \\
&               &       &           &            &           &            &            &          &           &           \\
\hline 
&               &       &           &            &           &            &            &          &           &           \\
J1555  & S\'{e}rsic  (B) & 4.16 & 1.39            & 0.51      & 17.99   & 17.51   & 1.79 & 0.64 & 0.36 & Lenticular (E)   \\
       & S\'{e}rsic  (D) & 0.50 & 4.49            & 0.39      & 18.62   &         &      &      &      &                  \\
J1601  & S\'{e}rsic      & 4.44 & 1.59            & 0.62      & 17.07   & 17.07   & -    & -    & -    & Elliptical (E)   \\
J1609  & S\'{e}rsic  (B) & 2.00 & 4.00            & 0.77      & 16.37   & 15.09   & 0.59 & 0.31 & 0.69 & Edge-on disk (L) \\
       & S\'{e}rsic  (D) & 1.50 & 3.97            & 0.04      & 17.00   &         &      &      &      &                  \\
       & S\'{e}rsic  (D) & 0.50 & 9.87            & 0.13      & 15.79   &         &      &      &      &                  \\
J1622  & S\'{e}rsic  (D) & 0.50 & 4.64            & 0.30      & 17.81   & 17.23   & 0.09 & 0.09 & 0.91 & Spiral (L)       \\
       & S\'{e}rsic  (D) & 0.50 & 9.33            & 0.54      & 18.42   &         &      &      &      &                  \\
       & PSF (B)        & -    & -               & -         & 19.90   &         &      &      &      &                  \\
J1630  & S\'{e}rsic      & 4.10 & 1.63            & 0.88      & 16.93   & 16.93   & -    & -    & -    & Elliptical (E) \\
&               &       &           &            &           &            &            &          &           &           \\
\hline
\end{tabular}
\end{table*}

More quantitative characterisation of the light distributions of the host galaxies was performed using the two-dimensional fitting algorithm \texttt{GALFIT} \citep{peng02,peng10}, which allows for the fitting of multiple structural components to galaxy light profiles. Our main motivation for doing this was to better identify the morphological types of the host galaxies (i.e. early-type vs. late-type), and to assess the relative contributions of bulge and disk light. In addition, the residual images generated as part of this process were also useful for clarifying any interaction signatures associated with the host galaxies. A well-defined procedure was followed to improve the consistency of the fitting process, starting with models of minimal complexity and only moving towards more complicated models if completely necessary. 

In much of the previous literature, galaxy light distributions have been modelled using S\'{e}rsic profiles, and hence we chose to adopt this method during our modelling to allow for direct comparison to previous work. In \texttt{GALFIT}, the S\'{e}rsic profile is described by
\[ \scalebox{1.2}{$
\Sigma(r) = \Sigma_{e} \rm{exp}\left\lbrace -\kappa \left(\frac{r}{r_e}\right)^{\frac{1}{n}} - 1 \right\rbrace,
$} \]

\noindent where $\Sigma$(\emph{r}) is the pixel surface brightness of the galaxy light at radius \emph{r}, $\Sigma$(\emph{r$_{e}$}) is the value at $r_e$, the effective radius of the light profile, \emph{n} is the S\'{e}rsic index, and $\kappa$ is a constant coupled to \emph{n} \citep{peng02,peng10}.

Traditionally, S\'{e}rsic profiles with indices around $n=1$ (exponential) have been found to provide good fits for galaxy disks, and those with $n=4$ (de Vaucouleurs) for galaxy spheroids (galaxy bulges and elliptical galaxies).  Galaxies that appear to show both disk-like and bulge-like components have then been modelled with a combination of these two types of profile, in an attempt to separate the light profile characteristics for bulge-disk decomposition. Detailed bulge-disk decomposition of very nearby disk galaxies with various Hubble types (from S0 to Sc), however, suggests that classical-bulge-like components can have S\'{e}rsic indices as low as $n \sim 1.5-2$ at very high resolutions \citep{gh17}. In an attempt to be as inclusive as possible whilst still separating disk-like and classical-bulge-like components clearly, we chose to constrain the S\'{e}rsic indices for disk-like and classical-bulge-like components to the following ranges: $n = 0.5$ to $1.5$ for disk-like (late-type) components; and $n=2$ to $5$ for galaxy spheroid (early-type) components. As well as better distinguishing between disk-like and bulge-like components for determining the overall morphological class of the galaxy, this also had the advantage of minimising the effect of degeneracies in the models.

In addition to the S\'{e}rsic profiles, \texttt{GALFIT} also requires a representation of the point spread function (PSF) in the form of a FITS image. All modelling profiles are convolved with the PSF to account for the quality of the observational data. In some cases, an individual PSF component was also required in the models to account for strong, unresolved AGN emission at the centre of the galaxies. All PSFs were generated using the average light profile of non-saturated stars in the final coadded images for each of the targets. Two-dimensional surface profile plots of the light from these stars were visually inspected to ensure that they provided an adequate representation of the PSF at good signal-to-noise, considering both the measured seeing of the observations and the characteristic shape of the response from the CCDs. Following this, PSFs could be used in combination with S\'{e}rsic profiles for the modelling process.

The procedure used for the fitting involved multiple stages and terminated as soon as a suitable model was obtained. All of our models were produced using at least one S\'{e}rsic profile and used up to a maximum of three S\'{e}rsic profiles to avoid unwarranted complexity in the fitting process. Neighbouring stars and galaxies were modelled simultaneously with the targets to minimise their effect on accurately modelling the host galaxy light profiles. In all cases, pixels above 70 per cent of the saturation level were masked prior to fitting to avoid any issues introduced by the CCD response close to saturation. The process is outlined here:
\begin{enumerate}
\item \textbf{Single S\'{e}rsic profile $-$} Initially, an attempt was made at fitting a single S\'{e}rsic profile, with a S\'{e}rsic index constrained to the ranges outlined above for each `disk-like' and `bulge-like' profiles. Determining which of the two profiles provided the best fit gave a first impression of the morphological class of the host galaxy. It was found that this stage alone generated a sufficient model for 4 out of the 30 galaxies (13 per cent), with no additional component required.
\item \textbf{Two S\'{e}rsic profiles $-$} Failing success after stage (i), an attempt was made at fitting a model with two constrained S\'{e}rsic components, in the order: 1) one disk-like profile and one bulge-like profile (bulge-disk decomposition); 2) two disk-like or two bulge-like profiles (two-component late-type or early-type). This combination proved sufficient for 14 out of the 30 galaxies (47 per cent), and provided the majority of successful models.
\item \textbf{Two S\'{e}rsic profiles + PSF $-$} For 3 of the 30 galaxies (10 per cent), a PSF component was required in addition to the two S\'{e}rsic profiles to model strong, unresolved AGN emission. The PSF components were only accepted after the presence of strong emission lines had been confirmed from inspection of the SDSS spectrum.
\item \textbf{Three S\'{e}rsic profiles $-$} The most complex cases required a model with three constrained S\'{e}rsic components to produce a suitable fit. These were all decomposed into disk-like and bulge-like components, either as two disk-like profiles and one bulge-like profile, or two bulge-like profiles and one disk-like profile. This was necessary for 2 of the 30 galaxies (7 per cent). If a good fit was still not obtainable at this point, the fitting process was terminated to avoid unwarranted complexity in the models.
\end{enumerate}

In total, the overall light profiles of 23 (77 per cent) of the host galaxies were deemed to be recovered with reasonable accuracy using this process. The majority of models for the remaining seven objects failed due to issues related to the highly disturbed nature of the galaxy morphologies due to mergers (J0757, J1243, J1257, J1351). The unusual morphology of J1324 meant that a suitable fit could not be obtained before the models became too complex $-$ it is possible that this target has also been disturbed by an interaction, though the authors did not deem this clear enough for the galaxy to be classified as such. In the case of J0810, the presence of a nearby bright star caused any modelling attempts to produce an inaccurate representation of the galaxy light profile. Finally, a large proportion of the inner regions of the galaxy J1147 was saturated in the exposures, so any model fitting to this galaxy was deemed unreliable. 

In cases that required at least two S\'{e}rsic profiles for accurate modelling, the morphology and model parameters for the sub-components were inspected to search for distinct bulge-like and disk-like components. For J0911, the model components cannot be separated in this way. In this case, both components have S\'{e}rsic indices that lie in the range $n=1-2$, and their two-dimensional morphologies are intermediate between those of bulges and disks. Such components have been identified in previous studies \citep{flo04,ins10,fal14,urb19}, and we refer to these as `intermediate' (I) components. We also note that J1036 shows a highly detailed spiral arm, bulge and bar structure, along with signatures of past or on-going interactions. Detailed fitting of these structures was not performed, and, as such, the accuracy of the model for this galaxy is reduced compared to the others.  

The key parameters for each of the individual components of the GALFIT models are presented in Table~\ref{tab:galfit_info}. Where appropriate, the derived bulge-to-disk light ratios (B/D) and their respective contributions to the total light profile (B/T and D/T) are reported. These ratios were not calculated for models that produced good fits to the overall light profile but did not separate into bulge-like and disk-like components. It is found that the majority (8 out of 15) of galaxies for which both bulge and disk components were required were disk-dominated, having a median bulge-to-disk ratio of B/D $\sim 0.8$. All four single S\'{e}rsic models and three two S\'{e}rsic models (one also with a PSF component) required only early-type profiles. When added to the bulge-dominated galaxies, it is hence found that 14 out of the 23 galaxies that could be modelled (61 per cent) have early-type profiles, with 8 out of 23 (35 per cent) having late-type profiles (disk-dominated) and 1 out of 23 (4 per cent) having an intermediate profile. 

\begin{figure}
\centering
    \includegraphics[width=\columnwidth]{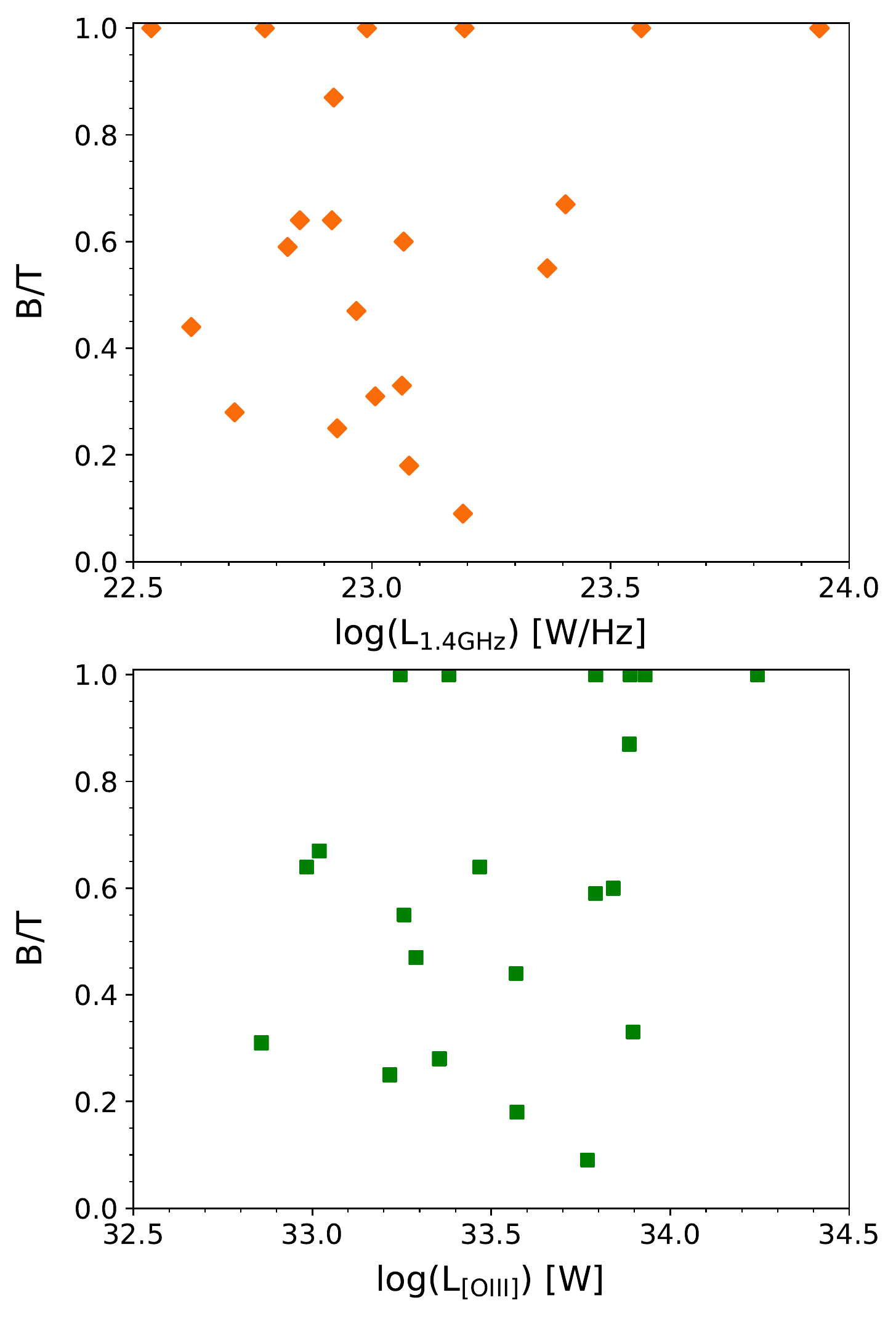}
    \caption{The fraction of the total light represented by the bulge-like components in the GALFIT models (B/T) against the radio powers (upper panel) and [OIII]\,$\lambda$5007 luminosites (lower panel) of the sources. All galaxies with successful models are included, excluding the galaxy that has only intermediate components (J0911). Galaxies with light profiles represented by only early-type components are considered to have B/T\,$=$\,1. }
   \label{fig:BT_both}
\end{figure}

Given that there is a mixture of morphological types found within the sample, it is also interesting to determine whether there is any relationship between the level of bulge-dominance in the galaxy light profiles and the AGN radio or optical luminosity. Figure \ref{fig:BT_both} shows the fraction of light contributed by the bulge-like components of the GALFIT models against both L$\rm_{1.4GHz}$ and L$\rm_{[OIII]}$ for all galaxies that were modelled successfully $-$ this excludes J0911, which has only intermediate components. Here, galaxies with only early-type components in their models are considered as having all of their light contributed by a bulge component. From this, it is clear that there is no trend in the bulge-dominance of the galaxy light with either L$\rm_{1.4GHz}$ or L$\rm_{[OIII]}$, as confirmed by Pearson correlation tests ($r=0.301$, $p=0.173$ and $r=0.338$, $p=0.124$, respectively). This suggests that although the radio-intermediate HERGs represent a greatly changed population of morphological types to their predominantly giant elliptical counterparts at high radio powers, the galaxy morphologies do not change gradually with the radio or optical luminosity of the AGN. Rather, there is a mix of morphological types at all optical/radio luminosities in the radio-intermediate range.

\section{Discussion}
\label{sec:disc}

Our analysis of deep \emph{r}-band images of 30 local radio-intermediate HERGs has allowed us to characterise their detailed optical morphologies, including both signatures of past or on-going interactions and their overall morphological types. In \S\ref{sec:prop}, we showed that 53 $\pm$ 9 per cent of the sample show interaction signatures, with the proportions within the most radio-powerful and least radio-powerful halves of the sample being 67 $\pm$ 12 per cent and 40 $\pm$ 13 per cent, respectively. These exact same proportions are found for the respective halves of the sample with the highest and lowest [OIII] luminosities. It is found, however, that all eight of the galaxies that show the highest levels of disturbance lie in the most radio-powerful half of the sample. This is not seen when the sample is divided in two by [OIII] luminosity, where 5 out of 8 these galaxies lie in the half of the sample with the highest [OIII] luminosities and 3 out of 8 in the half with the lowest [OIII] luminosities (see Figure \ref{fig:RPvsOIII}).

Both visual inspection and in-depth modelling of the host galaxy light distributions has revealed a mixture of morphological types within the sample (\S\ref{sec:vis} and \S\ref{sec:galfit}), with half showing evidence of a significant disk-like contribution to the galaxy light profile: 8 are classified as late-type based on \texttt{GALFIT} models (disk-dominated) and 14 as early-type (bulge-dominated or elliptical). The remainder proved unsuitable for accurate bulge-disk decomposition, four because of very highly disturbed morphologies resulting from mergers or interactions. Through comparison with previous work on radio AGN triggering and host morphologies, the implications of these results are discussed in this section.

\subsection{Triggering of the nuclear activity}
\label{sec:disc_trig}

\subsubsection{Importance of galaxy mergers and interactions}
\label{sec:all_int}

The focus of previous studies of radio AGN triggering has primarily been the most radio-powerful AGNs (e.g. those in the 3CR and 2 Jy samples), most probably because these were the easiest objects to study when the field began to develop. As discussed in \S\ref{sec:int}, radio AGNs with high-excitation emission spectra (HERGs or SLRGs), such as those considered here, are often linked with an accretion mode in which cold gas fuels the black hole through a geometrically thin, optically thick accretion disk \citep[e.g. see][]{hb14}. Study of exactly what would cause the radial inflow of gas towards the innermost regions to trigger the nuclear activity has produced mixed results, with possible mechanisms including the tidal torques introduced by galaxy mergers and interactions or the action of non-axisymmetric features in galaxy disks such as bars, oval distortions and spiral arms \citep[e.g.][]{hq10,hb14}.  

To date, the most extensive deep imaging study for radio-powerful AGNs with high-excitation optical emission has been carried out by \citet{ram11,ram12,ram13} for SLRGs in the 2 Jy sample ($0.05 < z < 0.7$), which provides strong evidence to suggest that major galaxy interactions could trigger the nuclear activity. \citet{ram11} find that 94 $\pm$ 4 per cent of the SLRGs in their sample show high-surface brightness signatures of past or on-going interactions, corroborating prior suggestions that radio galaxies with strong optical emission have highly disturbed optical morphologies \citep{heck86,sh89a,sh89b}. They also found that the rate of merger signatures was significantly higher than those found for control samples of elliptical galaxies in both the OBEY survey \citep{tal09} and the Extended Groth Strip \citep[EGS;][]{zhao09} matched in luminosity and redshift, when the same surface brightness limits are considered \citep{ram12}. In addition, \citet{ram13} find the 2 Jy SLRGs to preferentially reside in group-like environments that are denser than those of the EGS control galaxies but less dense than the cluster-like environments of the WLRGs in the sample. Group-like environments are well-suited to galaxy mergers, having a relatively high galaxy density without the high velocity dispersions of clusters that can suppress galaxy mergers \citep{quinn84,pb06}.

\begin{figure}
\centering
    \includegraphics[width=\columnwidth]{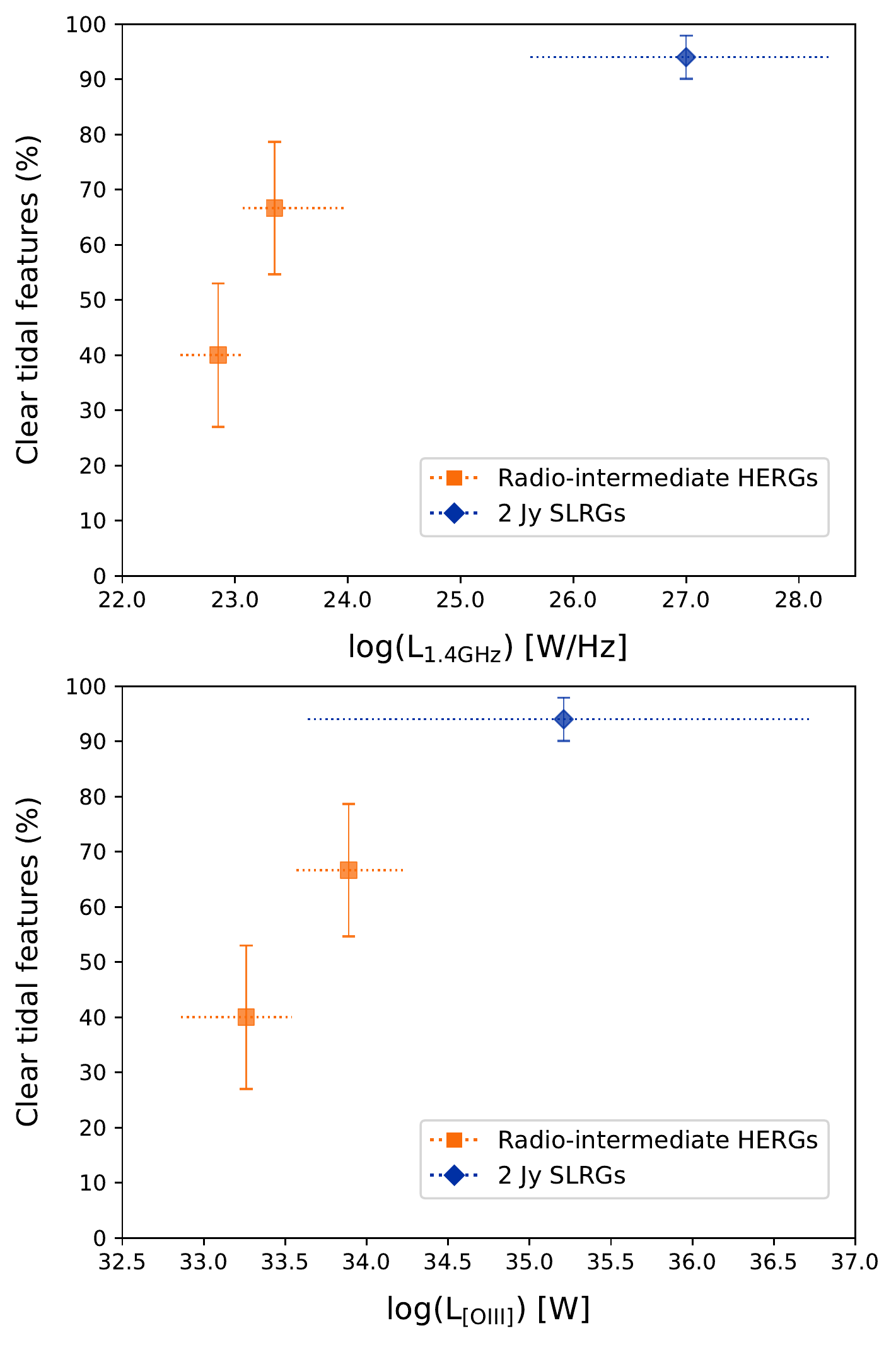}
    \caption{The percentage of HERGs in the current sample (orange squares) and SLRGs in the 2 Jy sample \citep[blue diamonds; ][]{ram11} that show clear tidal features, plotted against their radio powers at 1.4 GHz (top panel) and their [OIII]\,$\lambda$5007 luminosities (bottom panel). The results for each half of the radio-intermediate sample when split by their L$\rm_{1.4GHz}$ or L$\rm_{[OIII]}$ values are shown separately. The horizontal lines indicate the relevant L$\rm_{1.4GHz}$ or L$\rm_{[OIII]}$ ranges in each case, and the points represent their median values. Note that although the HERG and SLRG classifications are different, they are in broad agreement with one another.}
   \label{fig:TF_prop}
\end{figure}

In the upper panel of Figure~\ref{fig:TF_prop}, we plot the proportion of radio AGNs that show tidal features and other interaction signatures against radio power at 1.4\,GHz (L$\rm_{1.4GHz}$) for both our radio-intermediate HERGs and the SLRGs in the 2 Jy sample. Comparing the two samples in their entirety, the proportion of interacting galaxies in the radio-intermediate sample is found to be 53 $\pm$ 9 per cent (16/30), a difference at the 3.8\,$\sigma$ level from that found for the SLRGs in the 2 Jy sample \citep[94 $\pm$ 4 per cent; ][]{ram11}. Dividing our sample by radio power, it is also seen that disturbed galaxies are found more frequently in the most radio-powerful half (67 $\pm$ 12 per cent) of the sample than in the least radio-powerful half (40 $\pm$ 13 per cent) $-$ a difference at the 1.5\,$\sigma$ level. These results appear, therefore, to be consistent with a reduction in the proportion of radio AGNs with high-excitation optical emission lines that are triggered by galaxy interactions with decreasing radio power.

As shown in \S\ref{sec:prop}, however, the same proportions are found when the sample is divided into two equally populated bins in [OIII]\,$\rm \lambda$5007 luminosity (L$\rm_{[OIII]}$): 67 $\pm$ 12 per cent for the higher L$\rm_{[OIII]}$ half and 40 $\pm$ 13 per cent for the lower L$\rm_{[OIII]}$ half. The lack of correlation between L$\rm_{1.4GHz}$ and L$\rm_{[OIII]}$ for the galaxies in the sample also means that the galaxies giving the measured proportions in each case are, in fact, different (see Figure~\ref{fig:RPvsOIII}). In the lower panel of Figure~\ref{fig:TF_prop}, these proportions are again compared to those found for the 2 Jy SLRGs, but this time plotted against L$\rm_{[OIII]}$. From this, we see that the results are also consistent with a reduction in the level of merger-based triggering with decreasing optical (L$\rm_{[OIII]}$) power, often also considered to be a proxy for the bolometric luminosity of the AGN \citep[e.g.][]{heck04}. 

\begin{figure}
\centering
    \includegraphics[width=\columnwidth]{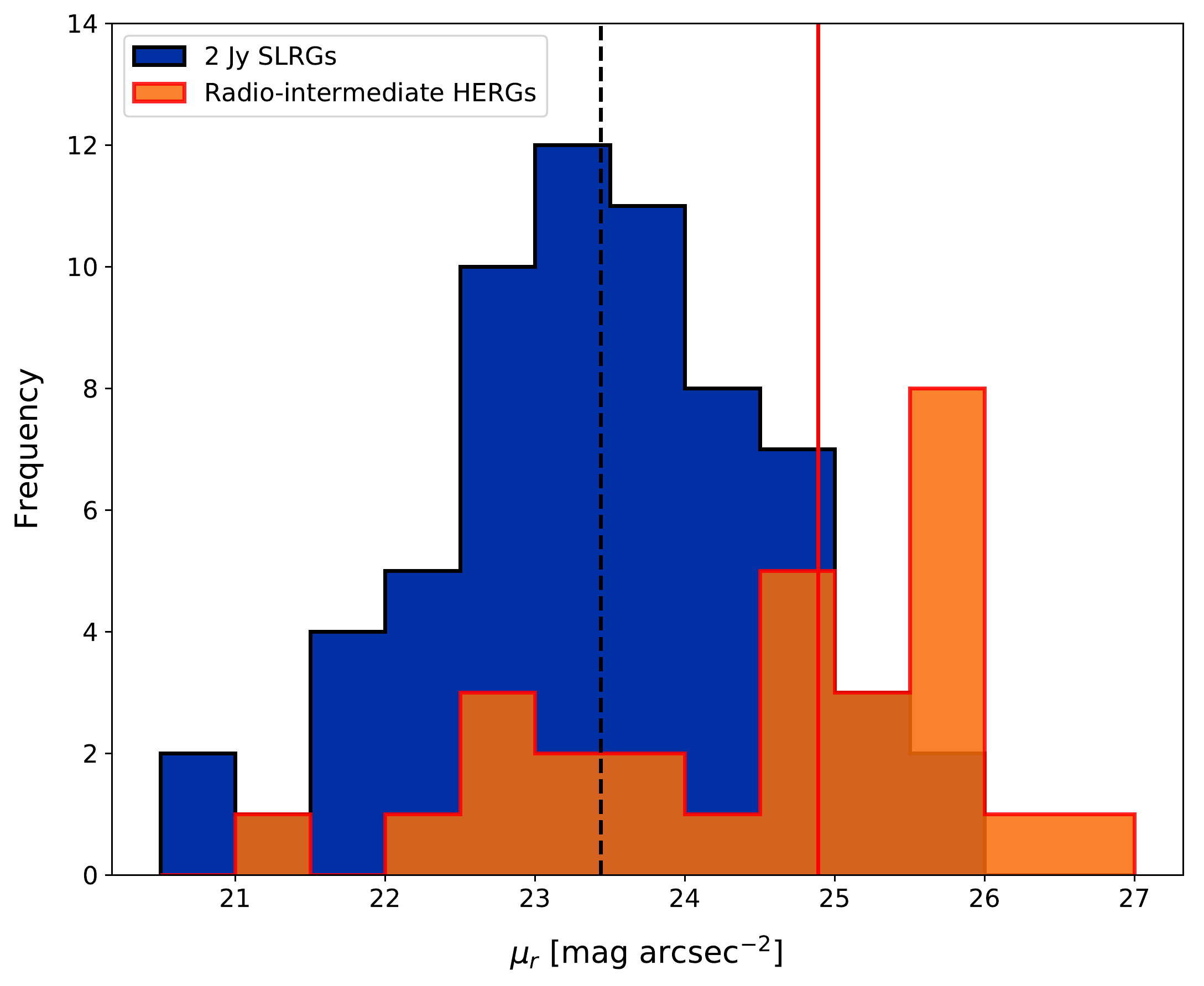}
    \caption{The distributions of tidal feature surface brightness measurements for the radio-intermediate HERGs (orange) and 2 Jy SLRGs \citep[blue;][]{ram11}. All values are in the AB system, and are corrected for surface brightness dimming, foreground galactic extinction and cosmological \emph{K}-correction. The median values are indicated by the solid red line for the radio-intermediate HERGs and the dashed black line for the 2 Jy SLRGs.}
   \label{fig:SB_dists}
\end{figure}

A further point of interest comes from comparing the measured surface brightnesses for the interaction signatures in the two studies. Figure~\ref{fig:SB_dists} shows the distribution of surface brightness measurements for the tidal features both from this work and \citet{ram11}, after accounting for surface brightness dimming, foreground galactic extinction and cosmological \emph{K}-correction. Although both studies have comparable image depths, we see that the 2 Jy SLRGs typically show interaction signatures of higher surface brightness $-$ the faintest securely detected features have (corrected) surface brightnesses of $\mu^{\rm{corr}}_r = $ 25.88 mag\,arcsec$^{-2}$ \citep{ram11} and $\mu^{\rm{corr}}_r = $ 25.99 mag\,arcsec$^{-2}$ (this work), but the median values are $\mu^{\rm{corr}}_r = $ 23.44 mag\,arcsec$^{-2}$ and $\mu^{\rm{corr}}_r = $ 24.84 mag\,arcsec$^{-2}$, respectively. This suggests that the galaxy interactions that trigger the 2 Jy SLRGs are likely to be more major mergers than those that may trigger the radio-intermediate HERGs.

We note, however, that there are some caveats to consider when comparing our results with those of the 2 Jy studies. Firstly, it is seen that all the SLRGs in the 2 Jy sample for which stellar masses have been measured have values $\sim 10^{11}$ $\rm M_{\odot}$ and above \citep{tad16}. This is above the median value for the radio-intermediate HERGs in the current sample (10$^{10.8}$ $\rm M_{\odot}$), which can have stellar masses that are down to an order a magnitude lower ($\sim$ 10$^{10}$ $\rm M_{\odot}$). In addition, a mixture of early-type and late-type galaxies are seen in the radio-intermediate HERG population, whereas the vast majority of 2 Jy radio galaxies are associated with early-type morphologies. Finally, although there is a high degree of overlap between the HERG and SLRG classifications, the schemes are not equivalent \citep[see][]{tad16}.

\subsubsection{Highly disturbed galaxies}
\label{sec:maj_int}
The discussion in the previous section was based on the detection of {\it any} sign of morphological disturbance, regardless of the level. This includes objects for which the low-surface-brightness features are only visible after careful manipulation of the image contrast levels. However, for a subset of eight objects, the signs of galaxy interactions are obvious even on cursory inspection of the images. This subset of highly disturbed systems includes all six of the galaxies that the authors deemed too disturbed to classify accurately by visual inspection (J0757, J0836, J0902, J1243, J1257, J1351; the `merger' class), four of which could also not be modelled with \texttt{GALFIT} on this basis (J0757, J1243, J1257, J1351). A further two galaxies are included based on the fact that they are strongly interacting with companion galaxies (J1358, J1412), as demonstrated by the presence of large-scale tidal features that provide an alternative indication of on-going major mergers \citep[e.g. see][]{duc15}. In addition, it is seen that 6 out of 8 of these galaxies have at least one tidal feature at relatively high surface brightness ($\leq 24$\,mag\,arcsec$^{-2}$), including all five galaxies with surface brightnesses greater than the median measured for the strongly disturbed 2 Jy SLRGs.

When the sample is divided in two by radio power, \emph{all} of these eight highly disturbed systems are found to lie in the half of the sample with the highest radio powers. In contrast, they lie in both halves of the sample when divided by their [OIII]\,$\lambda$5007 luminosities. Comparing the proportions found in each of the two halves in both cases, a 3.3\,$\sigma$ difference in proportions is found when dividing the sample by radio power, but only a 0.8\,$\sigma$ difference when dividing by [OIII]\,$\lambda$5007 luminosity. We again note that given that there are only eight of these highly disturbed galaxies in the entire sample, a more significant result could not have been observed when the division by radio power is made. Therefore, these resuts suggest that triggering through major galaxy interactions is linked with the radio powers of the HERGs, but not so strongly linked with their [OIII]\,$\lambda$5007 luminosities.

In combination with the results discussed in \S\ref{sec:all_int}, this is therefore consistent with a picture in which the relative importance of mergers and interactions for triggering radio AGNs with HERG spectra is related to their radio powers. In line with the results of \citet{ram11,ram12,ram13}, mergers appear to be important for triggering those with higher radio powers, with all of the most highly disturbed galaxies in the radio-intermediate sample lying in the half with the highest radio powers, notably including the three most-radio powerful cases. This is aligned with a reduction in the proportion of radio-intermediate HERGs that show any sign of disturbance from the most radio-powerful half to the least-radio powerful half of the sample, suggesting that other triggering mechanisms must be important when moving towards lower radio powers.

\subsection{Host morphologies}
\label{sec:disc_hosts}

As mentioned in \S\ref{sec:int}, powerful radio galaxies (typically L$\rm _{1.4GHz} > 10^{25}$ W\,Hz$^{-1}$) have long been associated with giant elliptical hosts \citep[e.g.][]{mms64,dun03,best05b}. There are, however, some suggestions that late-type hosts could be more common towards lower radio powers \citep[][]{sad14,tad16}. This could be particularly apparent for HERGs, if Seyfert galaxies are indeed considered as their low radio power equivalents. Crudely, one might therefore expect that radio-intermediate HERGs would manifest as a mixed population of galaxy morphologies, representing the transition from early-type galaxies at higher radio powers to late-type galaxies at lower radio powers. 

Table~\ref{tab:host_prop} compares the morphological classifications of the galaxies in the sample as obtained from visual inspection, both from this work and the Galaxy Zoo project, and the detailed light profile modelling performed using \texttt{GALFIT} $-$ the methods are described in \S\ref{sec:vis} and \S\ref{sec:galfit}, respectively. Regardless of the technique, it is immediately clear that there is no dominant class of host galaxy morphology, and that radio-intermediate HERGs do in fact represent a mixture of late-type and early-type morphologies. As discussed in \S\ref{sec:galfit}, however, we can also rule out the idea that the host morphology $-$ as quantified using the B/T ratio $-$ changes gradually with radio or optical AGN luminosity along the HERG sequence.

\begin{table}
\caption{The number and corresponding proportion of galaxies in the sample with the indicated morphological classfications. The results from visual inspection, both in this work and from the Galaxy Zoo project \citep{lin08,lin11}, and detailed light profile modelling using \texttt{GALFIT} \citep{peng02,peng10} are presented. Note that debiased Galaxy Zoo results were only available for 28 of the galaxies, as indicated. The quoted uncertainties are standard binomial errors.}
\label{tab:host_prop}
\begin{tabular}{m{4.1cm}C{1.5cm}C{1.5cm}}
\hline 
Visual inspection    & Number & Prop. \\ \hline
Late-type (Spiral/disk)    & 13                 & 43 $\pm$ 9 \%  \\
Early-type (Elliptical/Lenticular)    & 11                 & 37 $\pm$ 9 \%        \\
Mergers                     &   6        &      20 $\pm$ 7 \%    \\
\\ \hline
Galaxy Zoo                           &  Number & Prop. \\
\hline 
Late-type  (Spirals)                &        11          & 37 $\pm$ 9 \%        \\
Early-type (Ellipticals)             &       9                  &  30 $\pm$ 8 \%        \\
Uncertain (Merger/``Don't Know")                     & 10                  & 33 $\pm$ 9 \%        \\
\\ \hline
Galaxy Zoo $-$ \emph{debiased (28 only)}                             &  Number & Prop. \\
\hline 
Late-type  (Spirals)                &        12          & 43 $\pm$ 9 \%        \\
Early-type (Ellipticals)             &       7                  &  25 $\pm$ 8 \%        \\
Uncertain (Merger/``Don't Know")                     & 9                  & 32 $\pm$ 9 \%        \\
\\ \hline
Light profile modelling                 & Number & Prop.\\
\hline 
Late-type                     & 8                  &    27 $\pm$ 8 \%      \\
 $-$ \emph{one component}           &     \emph{0}                      &   \\
 $-$ \emph{disk-dominated}  &     \emph{8}                      &   \\
Early-type                   & 14                 &    47 $\pm$ 9 \%     \\
 $-$ \emph{one component}           &     \emph{7}                      &  \\ 
 $-$ \emph{bulge-dominated}           &     \emph{7}                      &  \\ 
Intermediate           &       1               &       3 $\pm$ 3 \% \\
Undetermined  & 7                  & 23 $\pm$ 8 \%        \\
 $-$ \emph{mergers (major disturbance)}           &     \emph{5}                      &   \emph{16 $\pm$ 7 \%}            \\
 $-$ \emph{fitting issues}           &     \emph{2}                      &   \emph{7 $\pm$ 5 \%}      \\

\end{tabular}
\end{table}

Considering the results of the visual inspection, we see that the proportion of galaxies classified as late-type and early-type are consistent between this work and Galaxy Zoo project, with a slight preference for late-types relative to early-types. The main difference between the two lies in the higher proportion of hosts categorised as ``Merger" or ``Don't Know" in the Galaxy Zoo results compared to those classified to be merging by the authors. This is perhaps unsurprising given the relative inexperience of the participants in the project and the common occurrence of lenticular galaxies, which are hard to categorise in the Galaxy Zoo classification scheme.

It is important to note, however, that visual classifications are always subject to biases \citep[e.g. see][]{cab18}, which, despite attempts \citep[e.g.][]{bam09,wil13}, are difficult to correct for, especially for small sample sizes. This provided strong motivation for performing detailed modelling of the galaxy light profiles, with the goal of providing a more quantitative basis for the galaxy classifications. Interestingly, the proportion of early-type hosts relative to late-types is seen to be higher in this case, in contrast with the results of the visual inspections. It is worth noting, on the other hand, that a total of 15 out of the 30 galaxies in the sample required both bulge-like and disk-like components to fit their light profiles, a proportion comparable to the number classified as late-types from the visual classification. In addition, the majority of multi-component models were disk-dominated, as is evidenced by the median bulge-to-disk ratio of B/D\,$\sim$\,0.8.

One further point of interest is that the presence of major disk components in the galaxy light profiles opens up the possibility of AGN triggering via secular mechanisms related to galaxy disks, as mentioned in \S\ref{sec:disc_trig}. Disk instabilities have been suggested as an important alternative triggering mechanism for AGNs below quasar luminosities \citep[e.g.][]{men14}, although it is uncertain whether this would be important outside of the violent conditions expected in clumpy disk galaxies at higher redshifts \citep{bou12}. However, it is thought that the weaker and flatter bulges sometimes seen in late-type galaxies could be built up by instabilities in the disks \citep{kk04}, which could simultaneously trigger the nuclear activity. In addition, bar-driven fuelling could contribute in at least some cases \citep[e.g.][]{gal15}. Given the reduced proportion of interacting galaxies and increased proportion of late-type hosts that are found in the sample relative to radio-powerful AGNs, this suggests that such processes could collectively provide a viable alternative triggering mechanism to mergers for radio-intermediate HERGs. 

\section{Summary and Conclusions}
\label{sec:conc}

Much previous work concerning radio AGNs has focused on those with the highest radio luminosities, due in most part to their ease of detection in the first flux-limited radio surveys. However, deeper surveys have revealed that radio-powerful examples are rare cases in the general population of local radio AGNs, and it has been shown that HERGs with intermediate radio powers can significantly disturb the warm and cool ISM phases of their hosts through jet-driven outflows. Therefore, the jets associated with radio-intermediate HERGs can potentially have a significant effect on the evolution of their host galaxies, and more detailed study is required to establish the properties of the population.

Our deep optical imaging study of a sample of 30 local radio-intermediate HERGs has allowed us to characterise the detailed optical morphologies of their host galaxies for the first time. Our main results are as follows:
\begin{itemize}
    \item  53 $\pm$ 9 per cent (16/30) of the galaxies in the sample show morphological signatures of a past or on-going merger event based on visual inspection by several independent classifiers, a significantly lower proportion (3.8\,$\sigma$) than for the SLRGs of the 2 Jy sample (94 $\pm$ 4 per cent).
    \item Proportionally, the most radio-powerful half of the sample has a 1.5\,$\sigma$ higher level of morphological disturbance than the least radio-powerful half $-$ 67 $\pm$ 12 per cent (10/15) and 40 $\pm$ 13 per cent (6/15), respectively. The exact same proportions are measured when dividing the sample by [OIII]\,$\lambda$5007 luminosity, however, despite no correlation being found between $\rm L_{1.4GHz]}$ and $\rm L_{[OIII]}$. 
    \item Eight galaxies in the sample have highly disturbed morphologies or large-scale tidal tails, suggestive of major galaxy interactions. It is found that all of these galaxies lie in the most radio-powerful half of the sample (a 3.3$\sigma$ result), including the three galaxies with the highest radio powers. In contrast, 5 out of 8 of these galaxies lie in the half of the sample with the highest [OIII]\,$\lambda$5007 luminosities and 3 out of 8 in the lower half, a difference of only 0.8$\sigma$.
    \item Results from detailed light profile modelling and visual classification indicate that the host galaxies of radio-intermediate AGNs have mixed morphological types. Many show signs of strong disk components (50 per cent; 15/30) in the models, and the majority of multi-component fits are disk-dominated (53 per cent; 8/15) with a median B/D $\sim$ 0.8. This contrasts with the predominantly early-type host galaxies of traditionally radio-powerful AGNs.
\end{itemize}

Taken together, these results suggest that the relative importance of triggering HERG activity through galaxy mergers and interactions reduces with radio power. However, mergers appear to remain important for triggering HERGs with higher radio powers, even within the radio-intermediate population. Moving to lower radio powers, the lower proportion of interaction signatures suggests a reduced importance for merger-based triggering. Given the higher proportion of late-type galaxy morphologies in the radio-intermediate population, this could be associated with an increased importance for secular triggering mechanisms related to galaxy disks (e.g. disk instabilities, the action of bars).

In future work we will expand this study to a larger sample of radio-intermediate HERGs, including objects at the higher end of the intermediate radio power range (L$\rm_{1.4 GHz}=10^{24}$$-$$10^{25}$ W\,Hz$^{-1}$). In this way we will further investigate the radio power dependence of the AGN triggering mechanism. This analysis will also be performed for a matched control sample of galaxies observed in the same way, to allow for direct comparison with non-active galaxies. In addition, we will use high-resolution VLA observations to determine the detailed radio properties of a subset of our sources, and hence greatly improve the general characterisation of this relatively unstudied population of radio AGNs.

\section*{Acknowledgements}
\label{sec:ack}

We wish to acknowledge Pierre-Alain Duc for his advice concerning parts of the image reduction, and for his assistance in interpreting the likely origins of the tidal features. We would also like to thank Katherine Inskip for providing the \texttt{IDL} script used to calculate the average PSF of the stars in the images. This research made use of the \emph{\emph{K}-corrections calculator} service available at http://kcor.sai.msu.ru/. Both JP and CT acknowledge support from STFC funding. CRA acknowledges the Ram\'{o}n y Cajal Program of the Spanish Ministry of Economy and Competitiveness through project RYC-2014-15779 and the Spanish Plan Nacional de Astronom\'{i}a y Astrof\'{i}sica under grant AYA2016-76682-C3-2-P. The Isaac Newton Telescope is operated on the island of La Palma by the Isaac Newton Group of Telescopes in the Spanish Observatorio del Roque de los Muchachos of the Instituto de Astrof\'{i}sica de Canarias. The WFC imaging observations were obtained as part of I/2017A/11.




\bibliographystyle{mnras}
\bibliography{master} 


\appendix

\section{Contamination by star forming galaxies}
\label{app:SFcont}
Given the moderate radio powers considered, the contamination of our radio-intermediate AGN sample by radio-quiet AGNs lying within star forming galaxies is possible, and must be considered. The level of this contamination is assessed here.

When generating their local radio AGN sample, \citet[][]{bh12} used the optical properties of the SDSS radio sources to perform three separate tests to distinguish between radio emission arising from star formation and from radio-loud AGN activity: i) the `D$_{4000}$ versus $L_{\rm rad}/M$' method of \citet{best05a}; ii) the `BPT' method of \citet{kau03}; iii) the `$L_{\rm H\alpha}$ versus $L_{\rm rad}$' method of \citet{kau08}. These tests are discussed in detail in their Appendix A, along with how their outcomes were used to determine whether the radio emission originated from star formation or AGN activity. We note that \textit{all} of the galaxies in our sample were deemed by \citet[][]{bh12} to have radio emission dominated by radio-loud AGN activity on this basis.

In addition, we perform an alternative test using the estimated star formation rates (SFRs) calculated for the 6 out of 30 galaxies (20\%) that are detected in at least one waveband as part of the Infrared Astronomical Satellite (IRAS) Faint Source Survey \citep[][]{mosh92}. The SFRs were calculated using the measured IRAS 60$\micron$ and 100$\micron$ fluxes, following the method of \citet[][]{cal10}. For the 2 out of these 6 galaxies not detected at 100$\micron$, the average 60$\micron$ to 100$\micron$ flux ratio for the other four galaxies was used in the calculations where necessary. The estimated radio emission at 1.4\,GHz resulting from galaxies with these SFRs was then derived using the prescription of \citet[][]{murph11}. From this, it is found that star formation could account for an average of only $\sim$20\% of the 1.4\,GHz luminosity obtained from the NVSS fluxes for the six galaxies considered $-$ for 5 out of the 6 galaxies, the contribution is $\lesssim$30\%, with the other (J0838) having the highest possible contribution, at $\sim$50\%. The results of this process are summarised in Table~\ref{tab:IR-radio}.

However, as shown by e.g. \citet[][]{mul11}, the infrared emission in the region of the spectrum covered by IRAS data could be subject to a significant contribution from AGN-heated dust emission, suggesting that the SFRs derived above are unreliable and should be treated as upper limits. In addition, we stress that 80\% of the galaxies in the sample do not have IRAS detections in the Faint Source Survey, and would thus have lower relative contributions towards the radio emission from star formation than those reported here. Given the low catalog detection rate and the likely overestimates of the 1.4\,GHz luminosities due to star formation for those that are detected, we argue that radio emission from radio-loud AGN activity provides the dominant contribution in all cases, supporting the results of the tests at optical wavelengths. Thus, it is deemed unlikely that there is significant contamination of our radio AGN sample by radio-quiet AGNs in star forming galaxies.

\begin{table}
\caption{Star formation properties for the six galaxies in the sample with far-infrared detections in the IRAS Faint Source Survey \citep{mosh92}. The target IDs and 60$\micron$ fluxes are presented, along with estimations of the star formation rates \citep[following][]{cal10}. The ratio of the expected resultant 1.4\,GHz luminosity \citep[following][]{murph11} to that derived from the NVSS fluxes is displayed in the final column.}
\label{tab:IR-radio}
\centering
\begin{tabular}{m{1.4cm}C{1cm}C{2.6cm}C{1.2cm}}
\hline \\
Target ID & F$_{\rm 60\micron}$ & Star formation rate &  \large{$\frac{\rm L^{\rm SF}_{\rm 1.4GHz}}{\rm L^{\rm NVSS}_{\rm 1.4GHz}}$}\\
    &    (Jy)   &   (M$_{\odot}$\,yr$^{-1}$)    & \\
 \\ \hline
                \\
J0810 & 0.685 & 33.2 & 0.10 \\
J0838 & 0.628 & 13.7 & 0.52 \\
J0950 & 0.410 & 5.94 & 0.04 \\
J1108 & 0.487 & 20.5 & 0.28 \\
J1243 & 0.449 & 29.0 & 0.20 \\
J1609 & 0.856 & 8.76 & 0.14 \\
\\
\hline
\end{tabular}
\end{table}


\bsp	
\label{lastpage}
\end{document}